\documentclass[11pt]{article}
\RequirePackage[T1]{fontenc}
\usepackage[english]{babel}
\RequirePackage{amsthm,amsmath}
\usepackage[a4paper]{geometry}
\usepackage{setspace}
\usepackage[usenames, dvipsnames]{color}
\usepackage{showlabels}
\usepackage{graphicx}
\usepackage{bm}
\usepackage{mathrsfs}
\usepackage[font={small,it}]{caption}
\usepackage{bbold}
\usepackage{multirow}
\usepackage{bbm}
\usepackage{chngpage}
\usepackage{adjustbox}
\usepackage{nicefrac}
\usepackage{amssymb}
\usepackage{bigints}
\usepackage{algorithm}
\usepackage{algpseudocode}
\usepackage{natbib}
\RequirePackage[colorlinks,citecolor=blue,urlcolor=blue]{hyperref}
\usepackage{xr}
\newcommand{\Real}{\mathbb{R}}
\newtheorem{theorem}{Theorem}[section]

\newtheorem{proposition}[theorem]{Proposition}

\newtheorem{definition}[theorem]{Definition}

\begin{document}
 \begin{center}

\textbf{\Large Testing One Hypothesis Multiple Times: \\The Multidimensional Case}
\vspace{1cm}

Sara Algeri$^1$ and David A. van Dyk$^2$\\
\vspace{0.5cm}
\textcolor{black}{
{\small $^1$ School of Statistics, University of Minnesota,  \\0461 Church St SE, Minneapolis, MN 55455, USA. \\Email: salgeri@umn.edu}}\\
{\small $^2$ Department of Mathematics, Imperial College London, \\180 Queen's Gate, London SW7 2AZ, UK}\\
 \end{center}
 \begin{center}
\small \section*{ Abstract}
The identification  of new rare signals  in  data, the detection of a sudden change in a trend, and the selection of competing models, are  among the most challenging problems in statistical practice. 
These challenges can be tackled using a test of hypothesis  where a nuisance parameter is present  only under the alternative, and a computationally efficient solution can be obtained by  the ``Testing One Hypothesis Multiple times'' (TOHM) method. In the one-dimensional setting, a fine discretization of the space of
the non-identifiable parameter is specified, and a global p-value is obtained by approximating the distribution of the supremum of the resulting stochastic process. In this paper, we propose a computationally efficient inferential tool to perform  TOHM in the multidimensional setting. Here, the approximations  of interest typically involve the expected Euler Characteristics (EC) of the excursion set of the underlying random field.  We introduce a simple  algorithm to compute the EC in multiple dimensions and for arbitrarily large significance levels. This leads to  an highly generalizable computational tool to perform hypothesis testing under non-standard regularity conditions. 
 \end{center}
\small{\textbf{Keywords: } Non-identifiability in hypothesis testing, Multidimensional signal search,  Non-nested models, Euler characteristics, Lipschitz-Killing curvatures, Graph Theory.}

\section{Introduction}
\label{intro}
\textcolor{black}{
In applied sciences, searches for new signals in data often reduce to a problem of detecting an unexpected mode, a variation in a trend, or a sudden change  in the association among the variables considered. From a statistical perspective, all these scenarios can be  characterized by a  structural change in the underlying model. }

\textcolor{black}{
The search for dark matter is one of  many examples that fall under this framework. Dark matter is the substance postulated in the 1930s by Jan Oort and Fritz Zwicky  \citep{o32, z33,  z37} to account for missing mass in the universe. Understanding its nature and proving experimentally its existence is  a hot topic in both particle physics and astronomy. 
One of the main physics collaborations focusing on the discovery of dark matter is the Fermi Large Area Telescope (LAT)  collaboration. The experiments conducted at Fermi LAT provide measurements of  photons emission over large regions of the sky with the goal of finding evidence for emission due to dark matter over  emission due to the cosmic background.}

\textcolor{black}{
To illustrate one of the many  statistical challenges which arise in this context, we consider a simplified example. Here the locations of photons emitted by the cosmic background are assumed to be uniformly distributed over the search region, whereas photon locations from the  dark matter source are assumed to be distributed as a bivariate Gaussian. Specifically, let $(x_1,y_1),\dots,(x_n,y_n)$, be the coordinates at which $n$ photons are observed, and assume that the pairs $(x_i,y_i)$ are i.i.d. realizations of a random vector $(X,Y)$ with density
\begin{equation}
\label{unif_gauss}
h(x,y|\theta_1,\theta_2)=(1-\eta)\frac{1}{\lambda(\bm{\Theta})}+ \eta\frac{1 }{k_{\theta_1\theta_2}}\exp\biggl\{-\frac{1}{2\nu^2}\biggl[(x-\theta_1)^2+(y-\theta_2)^2\biggl]\biggl\}
\end{equation}
where $\eta \in [0,1]$ is the relative intensity of the dark matter emission centered at $\bm{\theta}=(\theta_1,\theta_2)$, $\bm{\Theta}$ is the search region with area  $\lambda(\bm{\Theta})$,  and  $k_{\theta_1\theta_2}$ is a normalizing constant.  To reduce the computational cost of the simulations proposed in Section \ref{sim}, we assume that $\nu$ is a known constant and we fix it to 0.5.   While the model in \eqref{unif_gauss} represents a simplified scenario, it is straightforward to extend it to more realistic applications for dark matter searches \citep[e.g.,][]{refB3}.}

\textcolor{black}{
To assess if the signal of a dark matter source is present or not we test
 \begin{equation}
\label{test1}
H_0: \eta=0 \qquad \text{versus} \qquad H_1: \eta>0.
\end{equation}
Notice that under $H_0$,  $\bm{\theta}$  does not appear in   \eqref{unif_gauss} and is unidentifiable. Thus classical inferential procedures  \citep[e.g.,][]{wilks, chernoff} do not apply to the test in \eqref{test1}. The most common approach to address this issue is to conduct a Monte Carlo simulation of the distribution of the tests statistic considered. Alternatively, it is possible to test for the presence of a signal at each possible location over a fine grid  and then correcting for the multiplicity of  tests of hypothesis conducted \citep[see][for an extensive discussion on statistical methods used in the search of dark matter and their limitations]{conrad}. More recent work  in physics literature discuss  solutions based on random fields theory \citep{PL05,vitells}.}

\textcolor{black}{
Unfortunately, when dealing with stringent significance requirements,  as   is typically the case in the most crucial (astro)physics discoveries \citep{lyons2015}, deriving the distribution of the test statistics  via a Monte Carlo simulation or resampling methods \citep[e.g.,][]{boostrap} may be computationally prohibitive. This   is further aggravated when dealing with complex models for which even a single Monte Carlo replicate can be computationally expensive. Conversely, a multiple hypothesis testing approach may be of limited use because it may be overly conservative  \citep[e.g.,][]{bonferroni35, bonferroni36}, may inflate the probability of a type I error \citep[e.g.,][]{benjamini}, or require independence among the tests being conducted \citep[e.g.,][]{hochberg}. Finally, although solutions based on random fields appear promising, in their current formulation they require  substantial mathematical derivations of the main quantities involved \citep[e.g.,][]{PL05,PL06} or may be challenging to implement computationally in more than two dimension.
In this paper, we discuss  a simple solution, namely Testing One Hypothesis Multiple times (TOHM),  which  generalizes the methods proposed by \citet{gv10} and \citet{vitells}, and provide a novel computational strategy. The solution proposed in this manuscript  allows us to approximate small p-values while avoiding the need for case-by-case mathematical computations and  reduces drastically the number of Monte Carlo or bootstrap samples required. As additional advantage, the accuracy of the approximation proposed increases under stringent significance requirements;  thus, it is particularly well suited for astrophysical and other physics searches which typically impose stringent detection thresholds.}

\textbf{\emph{TOHM at a glance.}} \textcolor{black}{In general terms, the structural change that we aim to test for can be specified via a nuisance parameter, denoted by $\bm{\theta}$, which characterizes the alternative model but becomes meaningless under the null hypothesis. For instance, in the dark matter search in \eqref{unif_gauss}, the parameter  $\bm{\theta}$ characterizing the location of the signal has no meaning when the signal intensity, $\eta$, is zero. Thus, the problem is reduced to a test of hypothesis in presence of non-identifiability.} In this setting, the null hypothesis can be tested versus a sequence of \emph{sub-alternative hypotheses},  $H_{1}(\bm{\theta})$, one for each possible value of $\bm{\theta}$ over a fine grid. The observed \emph{sub-test statistics} are then combined into a \emph{global test statistic} from which the global p-value is obtained.  Hence, the name: \emph{Testing One Hypothesis Multiple times}. When $\theta$ is  one dimensional, this leads to a stochastic process indexed by $\theta$, and a global p-value is obtained by approximating the tail probability of the supremum of this process \citep[e.g.,][]{davies77,davies87}.
In \citet{TOHM}, the global p-value is efficiently computed by defining a simple expansion for the expectation of the number of upcrossings of the underlying process to bound the  tail probability of its supremum. The advantage of this expansion is that its leading term can be computed using a Monte Carlo simulation that is much smaller than the one required by a full simulation of the null distribution of the global test statistic.  
In addition to its computational advantages, \citet{TOHM} generalizes the approximation/bound of \citet{davies77,davies87} and \citet{gv10} for the Likelihood Ration Test (LRT), to the supremum of a wider class of stochastic processes. Like \citet{davies77,davies87}, however, \citet{TOHM} is limited to the case of $\theta$ being one-dimensional.

\textbf{\emph{TOHM and multiple hypothesis testing.}} In principle, the problem of detecting a structural change in  data can be formulated as a multiple hypothesis testing problem, where an ensemble of local p-values, one for each possible value of  $\bm{\theta}$ over a fine grid, is produced. The main goal  is to identify an adequate correction for the smallest of these p-values in order to guarantee the desired family-wise  probability of  type I error or rate of false discoveries. In TOHM, on the other hand,   an overall correction for the probability of type I error is generated intrinsically by exploring the topology of the stochastic process of interest to obtain the global p-value. 

\textbf{\emph{TOHM in multiple dimensions: framework and challenges.}} To perform TOHM in multiple dimensions  we rely on  fundamental results pertaining to the distribution of the suprema of random fields  \citep{worsley94,taylor2003,adlerbook, taylor2008}.
Specifically, we consider a random field indexed by the non-identifiable multidimensional parameter,  $\bm{\theta}$,  and we use the \emph{mean Euler characteristic (EC) of the excursion set} of the random field (to be introduced Section \ref{main}) to approximate the global p-value.  \textcolor{black}{As discussed in Section \ref{ECheur}, this approximation relies on the so called Euler Characteristic Heuristic and  thus we verify it numerically in our applied examples. }
Furthermore, closed-form expressions for the  expected EC typically depend on complicated functionals, such as the so-called Lipschitz-Killing curvatures (see Section \ref{main}), whose analytical form is often hard to derive explicitly.
Finally, numerical methods may be computationally challenging in multiple dimensions or when the threshold at which the excursion occurs is particularly high. Hence there is a need for  novel computational tools to adequately estimate these quantities.

\textbf{\emph{Main contributions of this paper.}} 
In order to overcome these  difficulties,   we develop a novel algorithm, based on  graph theory, to efficiently compute the EC  in multiple dimensions. The resulting outputs can then be used in a system of linear equations whose solution provides an estimate of the Lipschitz-Killing curvatures. The method proposed can
efficiently   perform bump-hunting in two or more dimensions and  tackle other problems where structural changes can be characterized by a multidimensional parameter (\textcolor{black}{e.g.,  the dark matter example above and  Examples  2 and 3 in Section \ref{main}}).  Additionally, from a theoretical perspective,  the ability to test when  a multidimensional parameter  is present only under the alternative further generalizes classical inferential procedures, such as the Likelihood Ratio Test, beyond  the standard regularity conditions including non-nested models comparisons  \citep{algeri16, TOHM} as shown in our Example 2.   \textcolor{black}{Finally, the \texttt{R} package \texttt{TOHM}, downloadable among the Supplementary Materials, aims to facilitate the implementation of TOHM in practical applications. }

The remainder of the paper is organized as follows. In Section \ref{main} we introduce the theoretical framework of TOHM in multiple dimensions. In Section  \ref{sim} we present both  a suite of simulation studies that validates the results of Section \ref{main}, and three applications of TOHM to real data in the context of bump-hunting, non-nested models comparison and break-point regression. A general discussion   appears in Section \ref{discussion}. \textcolor{black}{Proofs, regularity conditions and additional results are collected in Appendices \ref{regular} and \ref{ECapp}. Supplemental materials of this manuscript are 
available online. }

\textcolor{black}{
\section{TOHM in multiple dimensions}
\label{main}  
\subsection{Motivating Examples}
\label{examples}
Here we extend the results of \citet{TOHM} to the case where the data distribution under $H_1$ is
 characterized by a multidimensional parameter, $\bm{\theta}$, that is not identifiable under $H_0$. In addition to the dark matter search example introduced in Section \ref{intro}, hereafter referred to as Example 1, we consider the following two examples.}

{\bf\emph{Example 2: Non-nested model comparison.} } As discussed in \citet{algeri16} and  \citet{TOHM}, \textcolor{black}{in order to choose between two non-nested models $f(\bm{y},\bm{\gamma})$ and $g(\bm{y},\bm{\theta})$, we consider the comprehensive model
\begin{equation}
\label{mixture}
h(\bm{y},\eta,\bm{\gamma},\bm{\theta})=(1-\eta)f(\bm{y},\bm{\gamma})+\eta g(\bm{y},\bm{\theta})
\end{equation} 
with $\bm{y}\in \Real^q$, $\eta\in[0,1]$, $\bm{\gamma} \in \bm{\Gamma}\subseteq \Real^p$, $\bm{\theta} \in \bm{\Theta}$ and $\bm{\Theta}\subset \Real^D$}. We test both \eqref{test1}, and 
\begin{equation}
\label{eta1}
H_0: \eta=1 \qquad \text{versus} \qquad H_1: \eta<1.
\end{equation}
\noindent Specifically, suppose we aim to distinguish between a gamma and a log-normal distribution. Equation \eqref{mixture} becomes
\begin{equation}
\label{nonnest}
(1-\eta)\frac{e^{-y/\tau}y^{\gamma-1}}{k_{\tau\gamma}}+ \eta\frac{\exp\bigl\{-\frac{\ln y-\mu}{2\sigma^2}\bigl\} }{y k_{\mu \sigma}},
\end{equation}
where $\eta \in [0,1]$, $\gamma>0$, $\tau>0$, $k_{\tau\gamma}$ and $k_{\mu \sigma}$ are normalizing constants. In this case, the  parameter which is present only under the alternative is 
$\bm{\theta}=(\mu,\sigma)$ when testing \eqref{test1} and $\bm{\theta}=(\gamma,\tau)$ when testing \eqref{eta1}. 
The informative scenarios arising from \eqref{test1} and \eqref{eta1} are the following:
\begin{itemize}
\item if $H_0$ in \eqref{test1} is rejected and $H_0$ in \eqref{eta1} is not, the log-normal model is selected, 
\item if $H_0$ in \eqref{eta1} is rejected and $H_0$ in \eqref{test1} is not, the gamma model is selected. 
\end{itemize}
In all other cases \eqref{test1} and \eqref{eta1} are insufficient or inappropriate to select between the models being compared.

{\bf\emph{Example 3: Break-point regression with a change of trend.} } We consider a logistic-regression model where the presence of a break-point $\theta$ may introduce a polynomial  relationship between the logit of the probability of success and the explanatory variable $x$, i.e.,
\begin{equation}
\label{logistic}
\log\biggl(\frac{\pi_i}{1-\pi_i}\biggl)=\phi_1+\phi_2 x_i+\xi(x_i-\theta)^{\alpha}\mathbbm{1}_{\{ x_i\geq\theta \}} \quad \text{for } i=1,\dots,n
\end{equation} 
 where  $x_i \in \Real$ for all $i=1,\dots,n$ and are considered as fixed, $\mathbbm{1}_{\{ \cdot \}}$ is the indicator function,  $\bm{\theta}=(\theta,\alpha)$, \textcolor{black}{$\pi_i=P(Y_i=1)$, $Y_i\sim \text{Binomial}(m_i,\pi_i)$, and $m_i$ is the number of observations available for each value $x_i$.}
In this case, the test of hypothesis is
\begin{equation}
\label{test2}
H_0: \xi=0 \quad \mbox{versus} \quad H_1:\xi\neq0.
\end{equation}
\textcolor{black}{The goal of Section \ref{theory} is to establish a general framework to perform   tests of hypothesis like those in   \eqref{test1}, \eqref{eta1} or \eqref{test2}.}

\textcolor{black}{
\subsection{Theoretical framework}
\label{theory}
\textcolor{black}{To formalize the general setting,  consider a random sample $\bm{y}=(\bm{y}_1,\dots,\bm{y}_n)$, with independent components distributed as the random variables or random vectors $\bm{Y}_i$. Let $\bm{\theta} \in \bm{\Theta}\subset \Real^D$, with $D \geq 1$,  and  suppose that
for all $\bm{\theta} \in \bm{\Theta}$, it is possible to specify a  sub-test statistic, $W_n(\bm{\theta})$, which is a function of the $\bm{Y}_i$}, and whose asymptotic or exact distribution under $H_0$  is known to be the same as  some statistic $W(\bm{\theta})$, with known distribution. Similarly, letting $\bm{\theta}$  vary, we can consider a  $D$-dimensional random field indexed by $\bm{\theta}$, namely
$\{W_n(\bm{\theta})\}=\{W_n(\bm{\theta}), \bm{\theta} \in \bm{\Theta}\}$, whose exact or asymptotic distribution under $H_0$ is known to be the same as a random field $\{W(\bm{\theta})\}=\{W(\bm{\theta}), \bm{\theta} \in \bm{\Theta}\}$. 
We define the \emph{global test statistics} to be $\sup_{\bm{\theta} \in \bm{\Theta}}\{W_n(\bm{\theta})\}$  which, by the continuous mapping theorem, follows  the same  distribution as $\sup_{\bm{\theta} \in \bm{\Theta}}\{W(\bm{\theta})\}$ (exactly or asymptotically). To perform tests of hypothesis such as those in  \eqref{test1}, \eqref{eta1} or \eqref{test2} we consider  the global p-value  
\begin{equation}
\label{globalpD}
P\biggl(\sup_{\bm{\theta} \in \bm{\Theta}} \{W(\bm{\theta})\}>c\biggl),\qquad c \in \Real,
\end{equation}
where $c$ is the observed value of $\sup_{\bm{\theta} \in \bm{\Theta}}\{W_n(\bm{\theta})\}$.
}

In Examples 1 and 2, we choose $W_n(\bm{\theta})$  to be the LRT statistic, \textcolor{black}{i.e., $W_n(\bm{\theta})$ is the difference of the log-likelihood under $H_1$ and $H_0$ multiplied by a factor of two and evaluated at the Maximum Likelihood Estimates (MLEs) of the unknown parameters. }  The models involved in both Examples 1 and 2 are special cases of  \eqref{mixture}. Hence, the asymptotic distribution of $\{W_n(\bm{\theta})\}$ and its components can be derived on the basis of existing results in literature  for mixture models.  
Specifically, for each value of $\bm{\theta}$ fixed, \eqref{mixture} is characterized by $\bm{\gamma}\in \bm{\Gamma}\subseteq \Real^{p}$ and the one dimensional parameter $\eta$, which is tested on the boundary of its parameter space. \citet[][Theorem 3, Case 5]{self} show that, under suitable regularity conditions, the asymptotic distribution of the LRT is given by 
$Z^2\mathbbm{1}_{\{ Z\geq 0\}}$,
which corresponds to a $\bar{\chi}^2_{01}$ random variable \citep[as defined in][]{shapiro,lin,takemura97} and distributed as a 50:50 mixture of $\chi^2_1$ and zero.

\textcolor{black}{The asymptotic joint distribution of  $\{W_n(\bm{\theta})\}$ can be specified following the approach of \citet{ghosh}, who derive the asymptotic distribution of the LRT for finite mixture models of the form \eqref{mixture} with $\bm{y}\in \Real$ and $\bm{\Theta}\subset\Real $. However, since in our setting both $\bm{y}$ and $\bm{\theta}$  are allowed to be multidimensional,  in Appendix \ref{regular} we re-state the  regularity conditions  of \citet{ghosh} accordingly. These assumptions allow us to establish the following result.}
\textcolor{black}{\begin{proposition}
\label{prop1}
Consider the mixture model in \eqref{mixture}, for which we test either \eqref{test1} or \eqref{eta1}. 
If assumptions A0-A5 in Appendix \ref{regular} hold, under $H_0$, the LRT random field $\{W_n(\bm{\theta})\}$  converges  to
\begin{equation}
\label{ZconeD}
\{W(\bm{\theta})\}=\{Z(\bm{\theta})\}^2\mathbbm{1}_{\{Z(\bm{\theta})\geq\text{$0$} \}}\qquad\text{as $n\rightarrow\infty$,}
\end{equation}
where $\{Z(\bm{\theta})\}$  is  a Gaussian random field with mean zero, unit variance and covariance function depending on $\bm{\theta}$.
\end{proposition}}
\textcolor{black}{Equation \eqref{ZconeD} implies that $\{W(\bm{\theta})\}$ is distributed as a $\bar{\chi}_{01}^2$ random field \citep[as defined in][Remark 2]{taylor13}, 
i.e., a ``patchwork'' of a $\chi^2_1$ random field and a random field which is zero everywhere, with  components marginally distributed as  $\bar{\chi}^2_{01}$ random variables. In the Supplementary Material  we assess the  validity of assumptions A0-A5  for   Examples 1 and 2 which guarantee the applicability of Proposition \ref{prop1} (see Appendix \ref{regular} for the proof).}

\textcolor{black}{In  Example 3, we let $W_n(\bm{\theta})$ to be the signed-root-LRT, i.e.,
\begin{equation}
\label{gLRT}
W_n(\bm{\theta})=\text{sign}(\widehat{\xi})\sqrt{LRT_n(\bm{\theta}))}
\end{equation}
where $\widehat{\xi}$  is the MLE of $\xi$, and $LRT_n(\bm{\theta})$ is the LRT statistic evaluated at $\bm{\theta}$.
In   \citet{moran70}  \citep[see also][]{davies77}, the signed-root-LRT test statistic is shown to be equivalent to the normalized score function in \eqref{gLRT}, and  asymptotically normally distributed with  mean-zero and unit variance. Therefore, it is sufficient to show the asymptotic normality of the normalized score random field indexed by
$\bm{\theta}=(\theta,\alpha)$ in order to guaranteed that, for large samples, $\{W_n(\bm{\theta})\}$ is also  a Gaussian random field. Proposition \ref{prop2} establishes this result for Example 3 (see Appendix \ref{regular} for the proof).}

\textcolor{black}{\begin{proposition}
\label{prop2}
Consider the  model in \eqref{logistic} under which we test \eqref{test2}, and assume that the classical Cramer's conditions which guarantee normality and  consistency of the MLE \citep[][p.500]{cramer2} hold. Under $H_0$, the  random field $\{W_n(\bm{\theta})\}$, with components defined as in \eqref{gLRT}, converges to
\begin{equation}
\label{gauss}
\{W(\bm{\theta})\}=\{Z(\bm{\theta})\} \qquad\text{as $m_i\rightarrow\infty$ for all $i=1,\dots,n$,}
\end{equation}
where $\{Z(\bm{\theta})\}$  is  a Gaussian random field with mean zero, unit variance and covariance function depending on $\bm{\theta}$.
\end{proposition}}

\textcolor{black}{As discussed in Section \ref{goodness}, several challenges arise from a theoretical perspective in Example 3 due to the lack of smoothness of $\{Z(\bm{\theta})\}$ in \eqref{gauss} and necessary to approximate \eqref{globalpD} (see Section \ref{approx_pvals}). Numerically,  we find that   the approximation obtained for the global p-value is less accurate than in Examples 1-2 and leads to an upper bound for \eqref{globalpD}. }

Although we focus on test statistics based on the LRT, our method  can in principle be applied to any test statistic whose asymptotic or exact distribution is known. For instance, one may consider, among others, the  normalized Score, Lagrange Multipliers or Wald test statistics. While  these choices circumvent  the computational burden of the optimization  involved in the LRT, they require the calculation of the Fisher Information matrix and the covariance function of the resulting processes. This may introduce a substantial level of computational complexity when  the integrals involved can only be computed numerically. Additionally, even when the sample size is   moderately large,  the asymptotic  distribution may not be achieved  (see for instance \citet{algeri16} where a realistic dark matter search is conducted using only 200 events.) These concerns are carefully investigated via a suite of numerical studies in \citet{algeri2}.

\subsection{Approximating global p-values}
\label{approx_pvals}
In the one-dimensional setting,  \eqref{globalpD} is equivalent to the probability of observing at least one upcrossing of $\{W(\theta)\}$ above $c$. 
Specifically, we say that the process  $\{W(\theta)\}$  has an \emph{upcrossing} of a threshold $c\in \Real$ at $\theta_0 \in \Theta \subseteq \Real$ if, for some $\epsilon>0$, $\{W(\theta)\}\leq c$ in the interval $(\theta_0-\epsilon,\theta_0)$ and $\{W(\theta)\}\geq c$ in the interval $[\theta_0,\theta_0+\epsilon)$ \citep[e.g.][]{adler2000}.
 This definition however, is unhelpful in the multidimensional setting. Therefore, our first aim is to identify a generalization of the number of upcrossings  in the context of random fields. 
\textcolor{black}{This can be done by means of a heuristic argument known as \emph{Euler Characteristic Heuristic}  \citep{adler81, adler2000}; this is described in Section \ref{ECheur}.  }

\begin{figure}
\centering
\includegraphics[width=13cm]{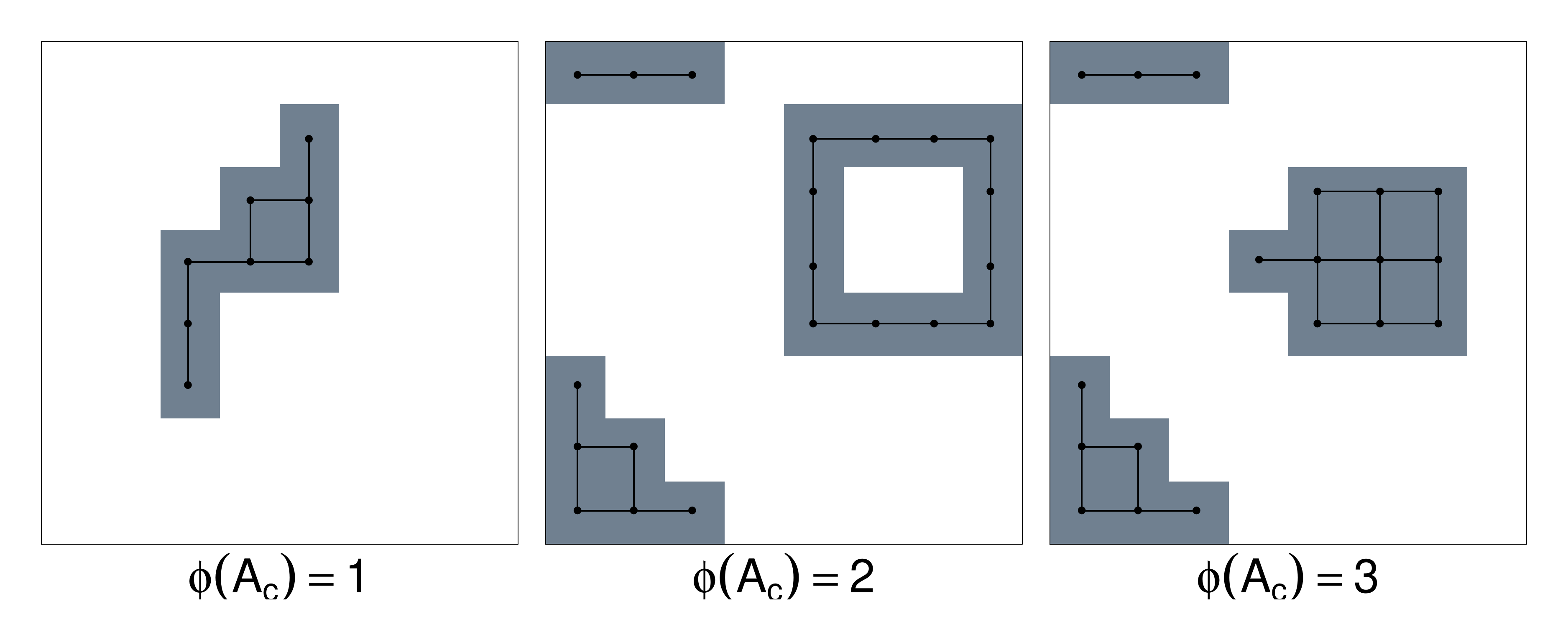}
\caption{The shaded regions illustrate three possible excursion sets $\mathcal{A}_c$. The Euler characteristic (EC) of $\mathcal{A}_c$ in the left, central and right panels are 1,2 and 3, respectively. The EC can be obtained by counting the number of connected components less the number of holes of $\mathcal{A}_c$. Alternatively,  considering a quadrilateral mesh of the image (black points and black edges), the same EC is  given by  the number of points less the number of edges plus the number of faces (squares).}
\label{eulerfig}
\end{figure}

\textcolor{black}{
\subsubsection{The Euler Characteristic Heuristic} 
\label{ECheur}
 \citet{hasofer78} noted that the relationship between \eqref{globalpD} and the probability of an upcrossing can be extended to
 multiple dimensions} by considering
the number of local maxima\footnote{We are interested in scenarios where local maxima become rarer and rarer as $c\rightarrow +\infty$. Hence, we are implicitly assuming that no ridges above  $c$ occur. } of $\{W(\bm{\theta})\}$ that exceed $c$, namely $M_c$, hence
\begin{equation}
\label{localMax}
P\biggl(\sup_{\bm{\theta} \in \bm{\Theta}} \{W(\bm{\theta})\}>c\biggl)=P(M_c\geq1)\leq E(M_c).
\end{equation}
where the inequality follows from Markov's inequality.
Unfortunately,  analytical expressions for  $E(M_c)$ are known only asymptotically in $c$, and thus  cannot be exploited to derive multidimensional counterpart of  \citet{TOHM},   which rely on evaluating $E(M_{c_0})$ at an arbitrarily small $c_0$.
 A quantity that is more amenable and  for which analytical expressions are known exactly, is   the \emph{expected  Euler characteristic (EC)  of the excursion set of $\{W(\bm{\theta})\}$ above $c$}. A clear  description of the EC requires a few concepts from  geometry that we now summarize \citep[see][]{adler2000}.
\begin{definition} 
\label{excursion}
The excursion set of $\{W(\bm{\theta})\}$ above $c$ is the set of points
\begin{equation}
\label{excSet}
\mathcal{A}_c=\{\bm{\theta} \in \bm{\Theta}:W(\bm{\theta})\geq c\}.
\end{equation}
\end{definition}
\begin{definition} 
\label{excursion}
The Euler characteristic, $\phi(A)$, of a compact set  $A\subset \Real^D$ is the additive, integer-valued functional of $A$ uniquely determined by the following properties:
\begin{equation}
\phi(A) = \left\{ \begin{array}{ll}
        1 & \mbox{if $A$ is homeomorphic to a $D$-dimensional sphere};\\
        0 & \mbox{otherwise}\end{array} \right. 
\end{equation}
\textcolor{black}{i.e., there exists a continuous mapping between $A$ and the $D-$dimensional sphere, whose inverse is also continuous,}
and
\[
\phi(A\cup B)=\phi(A)+\phi(B)-\phi(A\cap B).
\]
\end{definition}
\noindent Intuitively, in two dimensions  the EC of $\mathcal{A}_c$ is the number of  its connected components less its number of  ``holes'', see  Figure \ref{eulerfig}.  

\textcolor{black}{The heuristic argument of \citet[Chapter 6]{adler81} (see also \citet{adler2000}),  aims to approximate $P\bigl(\sup_{\bm{\theta} \in \bm{\Theta}} \{W(\bm{\theta})\}>c\bigl)$ using the expected EC. Specifically, \citet{adler81} notices that  the maxima of $\{W(\bm{\theta})\}$ above large values of $c$ can be approximated by  elliptic paraboloids, which correspond to simple connected components of $\mathcal{A}_c$ (in two dimensions for instance, ``simple'' refers to  components which are connected but do not contain any hole). Additionally,   as  $c\rightarrow\infty$,  the holes within the components of $\mathcal{A}_c$   disappear and 
EC approaches the number of simple connected components around each local maxima exceeding $c$.
It follows that the left hand side of the inequality in \eqref{localMax} can be approximated by the expected EC of $\mathcal{A}_c$, namely, $E[\phi(\mathcal{A}_c)]$, and thus we write 
\begin{equation}
\label{ECpval}
P\biggl(\sup_{\bm{\theta} \in \bm{\Theta}} \{W(\bm{\theta})\}>c\biggl)\approx E[\phi(\mathcal{A}_c)]\qquad\text{as $c\rightarrow\infty$},
\end{equation}
where  the $\approx$ sign  indicates that the   difference between the right and the left hand side approaches zero as $c\rightarrow\infty$ \citep[see][Chapter 14]{adlerbook}.  It is important to point out that, unlike $E[M_c]$, in principle $E[\phi(\mathcal{A}_c)]$ can be negative. This means that $E[\phi(\mathcal{A}_c)]$ does not bound the left hand side of \eqref{ECpval} from above.} 

\textcolor{black}{Unfortunately, since the approximation in \eqref{ECpval} is based on a purely heuristic argument, its  validity is not clear. Furthermore,  evaluating its accuracy   is a particularly challenging task, and  the error of the approximation is known only in a limited number of cases \citep[e.g.,][]{taylortakemura, taylor2008}. Since existing results cannot be easily extended to the applications considered in this manuscript, we rely on a numerical assessment of the accuracy of \eqref{ECpval} for Examples 1-3 by means of Monte Carlo simulations (see Section \ref{goodness} and Figure \ref{assess}).}

\vspace{-0.8cm}
\textcolor{black}{
\subsubsection{The Exepected Euler Characteristic}  
Another difficulty arising from the EC heuristic is the computation of $E[\phi(\mathcal{A}_c)]$.
\citet{worsley94, worsley95} and \citet{adler2000} among others, give analytical expressions of $E[\phi(\mathcal{A}_c)]$ for  isotropic random fields. 
The seminal work of   \citet{taylor2003}, \citet{adlerbook} and \citet{taylor2008}   generalizes these approaches
 to non-isotropic  random fields of arbitrarily large  dimension.  They provide a convenient expansion of $E[\phi(\mathcal{A}_c)]$   for  Gaussian-related (e.g., $\chi^2$, $t$, $F$, $\bar{\chi}^2$ etc) random fields. Specifically, if $\{W(\bm{\theta})\}$ is a real-valued random field that can be written as a function of  i.i.d.  mean zero, unit variance and suitably regular}\footnote{\textcolor{black}{A Gaussian random field $\{Z(\bm{\theta})\}$ is said to be ``suitably regular'' if it has almost surely  continuous partial derivatives up to the second order, and if the two-tensor field induced by $\{Z(\bm{\theta})\}$ satisfies the additional mild  conditions  specified in Definition 3.2 of  \citet{taylor2003}.}} \textcolor{black}{  Gaussian random fields, its expected EC can be written as}
\begin{equation}
\label{lipscitz}
 E[\phi(\mathcal{A}_c)]=\sum^{D}_{d=0}\mathcal{L}_d(\bm{\Theta})\rho_d(c),
\end{equation}
where  the $\rho_d(c)$ are functionals known as EC densities, and only depend on the (identical) marginal distribution of each component of $\{W(\bm{\theta})\}$ (see Appendix \ref{ECapp}). For example, $\rho_0(c)=P(W(\bm{\theta})>c)$. Closed-form expressions of  $\rho_d(c)$ are available in literature for  Gaussian, $\chi^2$, $F$ and  other Gaussian-related random fields \citep{taylor2003, adlerbook, taylor2008}. The functionals $\mathcal{L}_d(\bm{\Theta})$ are known as the  Lipschitz-Killing curvatures of $\bm{\Theta}$. Intuitively, they measure the intrinsic volume of   $\bm{\Theta}$, i.e., they account for its volume, surface  area, and boundaries. Their analytical forms typically rely on the covariance structure and partial derivatives of $\{W(\bm{\theta})\}$. 

 Unfortunately, obtaining closed-form expressions for  $\mathcal{L}_d(\bm{\Theta})$  is    challenging for non-isotropic fields \citep{adlerbook}. Even in the isotropic case this may require tedious calculations and  knowledge of the distribution of the derivatives of $\{W(\bm{\theta})\}$. In   Sections \ref{TOHMmultiD} and \ref{graphs} we introduce a simple approach to estimate the  $\mathcal{L}_d(\bm{\Theta})$  in \eqref{lipscitz}, and consequently, to compute the approximation of the global p-value in \eqref{ECpval}.

\subsection{Methodological setup}
\label{TOHMmultiD}
In this section we extend the results of \citet{TOHM} with the goal of efficiently computing the right hand side of \eqref{lipscitz}. 

\textcolor{black}{This can be done following the approach of \citet{vitells} in  two dimensions, 
and  further formalized  in  \citet{adler2017} in a multi-dimensional setting. Specifically, we consider a
sequence of constants $c_1\neq c_2\neq \dots\neq c_D$, with $c_k\in \Real$ for $k=1,\dots, D$. Notice that, under suitable smoothness conditions \citep[][see also Section \ref{goodness}]{taylor2003}, 
\eqref{lipscitz} holds for any  value $c$. Hence we can specify the following system of linear equations
\begin{equation}
\label{system}
\begin{cases}
E[\phi(\mathcal{A}_{c_1})]-{\mathcal{L}}_0(\bm{\Theta})\rho_0(c_1)&=\quad \sum^{D}_{d=1}\mathcal{L}_d(\bm{\Theta})\rho_d(c_1)\\
E[\phi(\mathcal{A}_{c_2})]-{\mathcal{L}}_0(\bm{\Theta})\rho_0(c_2) &=\quad \sum^{D}_{d=1}\mathcal{L}_d(\bm{\Theta})\rho_d(c_2)\\
& \vdots \\
E[\phi(\mathcal{A}_{c_D})]-{\mathcal{L}}_0(\bm{\Theta})\rho_0(c_D)&=\quad  \sum^{D}_{d=1}\mathcal{L}_d(\bm{\Theta})\rho_d(c_D),\\
\end{cases}
\end{equation}
where the $\mathcal{A}_{c_k}$ are  the  excursion sets of $\{W(\bm{\theta})\}$ above  the constants $c_k$ and  $E[\phi(\mathcal{A}_{c_k})]$ are the expected EC of $\mathcal{A}_{c_k}$.}

\textcolor{black}{In \eqref{system}, the Lipschitz-Killing curvature for $d=0$, ${\mathcal{L}}_0(\bm{\Theta})$, is known  and corresponds to the EC of $\bm{\Theta}$ \citep{taylor2008} (e.g., ${\mathcal{L}}_0(\bm{\Theta})$ is $0,1,1$ or $2$ if $\bm{\Theta}$ is a circle, a disc, a square or a cube, respectively). Thus, ${\mathcal{L}}_0(\bm{\Theta})$ need not to be estimated. Whereas, given the linear independence of the EC densities $\rho_d(\cdot)$ evaluated at different $c_k$, the 
solutions $\mathcal{L}^*_d(\bm{\Theta})$, $d=1,\dots,D$, of \eqref{system} provide expressions for the
Lipschitz-Killing curvatures for $d>0$. Hence, we can rewrite \eqref{lipscitz} as
  \begin{equation}
\label{expect2}
E[\phi(\mathcal{A}_c)]={\mathcal{L}}_0(\bm{\Theta})P(W(\bm{\theta})>c)+\sum^{D}_{d=1}{\mathcal{L}^*}_d(\bm{\Theta})\rho_d(c),
\end{equation}}

Finally, the approximation to the global p-value in \eqref{ECpval} can be restated, on the basis of \eqref{expect2}, as
\begin{equation}
\label{technical_bound3}
 P\biggl(\sup_{\bm{\theta} \in \bm{\Theta}} \{W(\bm{\theta})\}>c\biggl)\approx {\mathcal{L}}_0(\bm{\Theta})P(W(\bm{\theta})>c)+\sum^{D}_{j=1}\mathcal{L}^*_d(\bm{\Theta})\rho_d(c)
\end{equation}
\textcolor{black}{for large values of $c$.} Notice that when the $E[\phi(\mathcal{A}_{c_k})]$ are known, \eqref{expect2}
is an exact equivalence and thus the accuracy of the approximation in \eqref{technical_bound3} is the same as  \eqref{ECpval}.
 
In practice the $E[\phi(\mathcal{A}_{c_k})]$ are unknown and estimated  via a Monte Carlo simulation (details are given in Section~\ref{graphs}).
In Section \ref{sim} we discuss choices of the constants $c_k$ to reduce the computational time while preserving the accuracy of the Monte Carlo estimates of $E[\phi(\mathcal{A}_{c_k})]$.

\subsection{Computing the mean Euler characteristic via graphs}
\label{graphs}
\textcolor{black}{
To compute the Lipschitz-Killing curvatures ${\mathcal{L}^*}_d(\bm{\Theta})$ involved in the approximation of the global p-value in  \eqref{technical_bound3}, we estimate the quantities $E[\phi(\mathcal{A}_{c_k})]$, for $c_1,\dots,c_D$ via a Monte Carlo simulation. This requires the evaluation of $\phi(\mathcal{A}_{c_k})$ for a sequence of realizations of $\{W(\bm{\theta})\}$. In this section we propose a convenient algorithm to achieve this goal. }

To simplify notation, we assume that $\bm{\Theta}$ is the cross product of the parameter spaces of  components  $\bm{\theta}$. Specifically,
$\bm{\Theta}=\Theta_1\times\dots\times\Theta_D$, where $\Theta_d$ is the parameter space of  component $d$ of $\bm{\theta}$; the same reasoning easily applies when $\bm{\Theta}\subset \Theta_1\times\dots\times\Theta_D$ (e.g., Example 1 described in Section \ref{casestudies}). In practice, we can only evaluate $\{W(\bm{\theta})\}$ on a finite set of  values for $\bm{\theta}$. We do so by placing a grid of $R_d$ points on $\Theta_d$, for $d=1,\dots,D$ and evaluating $\{W(\bm{\theta})\}$ at $\bm{\theta}_r=(\theta_{r1},\dots,\theta_{rD})$ for $r=1,\dots,R$, with $R=R_1\times\dots\times R_D$, so that the evaluation points are the cross products of the component-wise grids.
Finally, we let $\tilde{\Theta}_d$ be the ordered set of evaluation points of   component $d$ of $\bm{\theta}$  and let $\bm{\Theta}_\times$ be the full set of evaluation points of $\bm{\theta}$ over the cross product of $\tilde{\Theta}_1,\dots,\tilde{\Theta}_D$, i.e., $\bm{\Theta}_\times=\{\bm{\theta}_r,r=1,\dots,R\}$.
For each constant $c_k$ in \eqref{system}, we define the excursion sets of $\{W(\bm{\theta}_r)\}$ above  $c_k$ to be the set of evaluation points   $\tilde{\mathcal{A}}_{c_k}=\{\bm{\theta}_r \in \bm{\Theta}_\times: W(\bm{\theta}_r)\geq c_k\}$, hence $\tilde{\mathcal{A}}_{c_k}\subseteq \bm{\Theta}_\times$ provides a discretization of $\mathcal{A}_{c_k}$. 
In order to compute $\phi( {\mathcal{A}}_{c_k})$ numerically, we consider a quadrilateral mesh of $ {\mathcal{A}}_{c_k}$ \citep{taylor2008}, i.e., the set of vertices composed of the points in $\tilde{\mathcal{A}}_{c_k}$ and the edges that connect them to form a partition of ${\mathcal{A}}_{c_k}$ into $D$-dimensional hyperrectangles, and  denoted   by $\mathcal{M}_k$.
Specifically, we consider the  set of edges, $E^1_{k}$, such that two vertices $\bm{\theta}_r$ and $\bm{\theta}_s$ in $\tilde{\mathcal{A}}_{c_k}$  are joined by an edge if and only if
\begin{equation}
\label{distance}
\mathtt{d}_{\varphi}(\bm{\theta}_r,\bm{\theta}_s)=\sqrt{\sum^{D}_{d=1}(\varphi_d{(r)}-\varphi_d{(s)})^2}=1,
\end{equation}
where, $\varphi_d{(r)}$ is the index of  component  $d$ of $\bm{\theta}_r$ within its (ordered) grid of evaluation points $\tilde{\Theta}_d$
  and $\mathtt{d}_{\varphi}(\bm{\theta}_r,\bm{\theta}_s)$ is the  Euclidean distance between the $D$ indexes of the $D$ components of $\bm{\theta}_r$ and $\bm{\theta}_s$ within the component-wise grids $\tilde{\Theta}_1,\dots,\tilde{\Theta}_D$.
In $\mathcal{M}_k$,  the lengths of the edges in $E^1_k$ are  the Euclidean distances between $\bm{\theta}_r$ and $\bm{\theta}_s$, i.e., $\mathtt{d}(\bm{\theta}_r,\bm{\theta}_s)=\sqrt{\sum^{D}_{d=1}(\theta_{rd}-\theta_{sd})^2}$. In quadrilateral meshes involving only unit hypercubes $\mathtt{d}(\bm{\theta}_r,\bm{\theta}_s)=\mathtt{d}_{\varphi}(\bm{\theta}_r,\bm{\theta}_s)$. 

\begin{figure}
\centering
\begin{adjustwidth}{0cm}{0cm}
\begin{tabular*}{\textwidth}{@{\extracolsep{\fill}}@{}c@{}c@{}c@{}}
      \includegraphics[width=50mm]{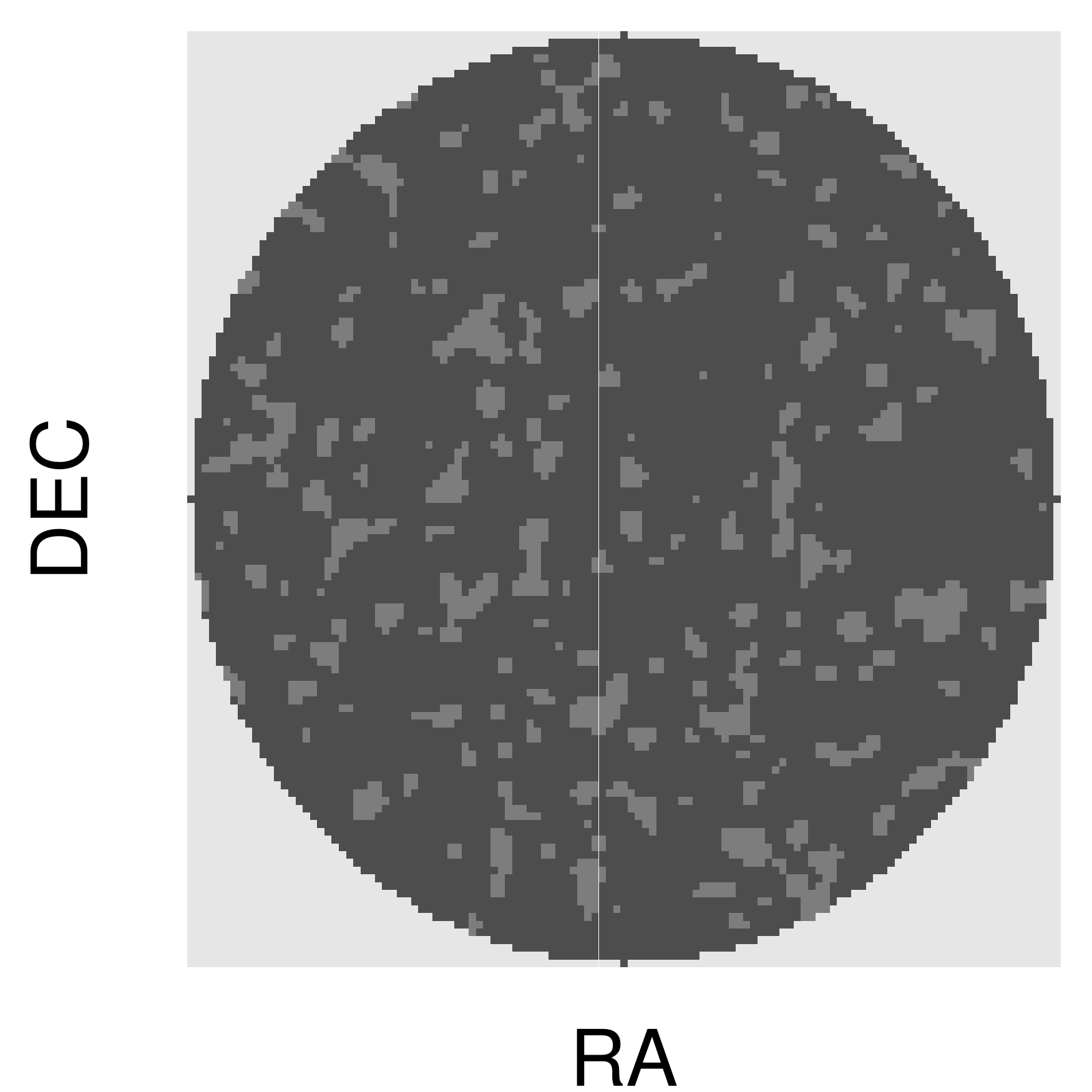} & \includegraphics[width=50mm]{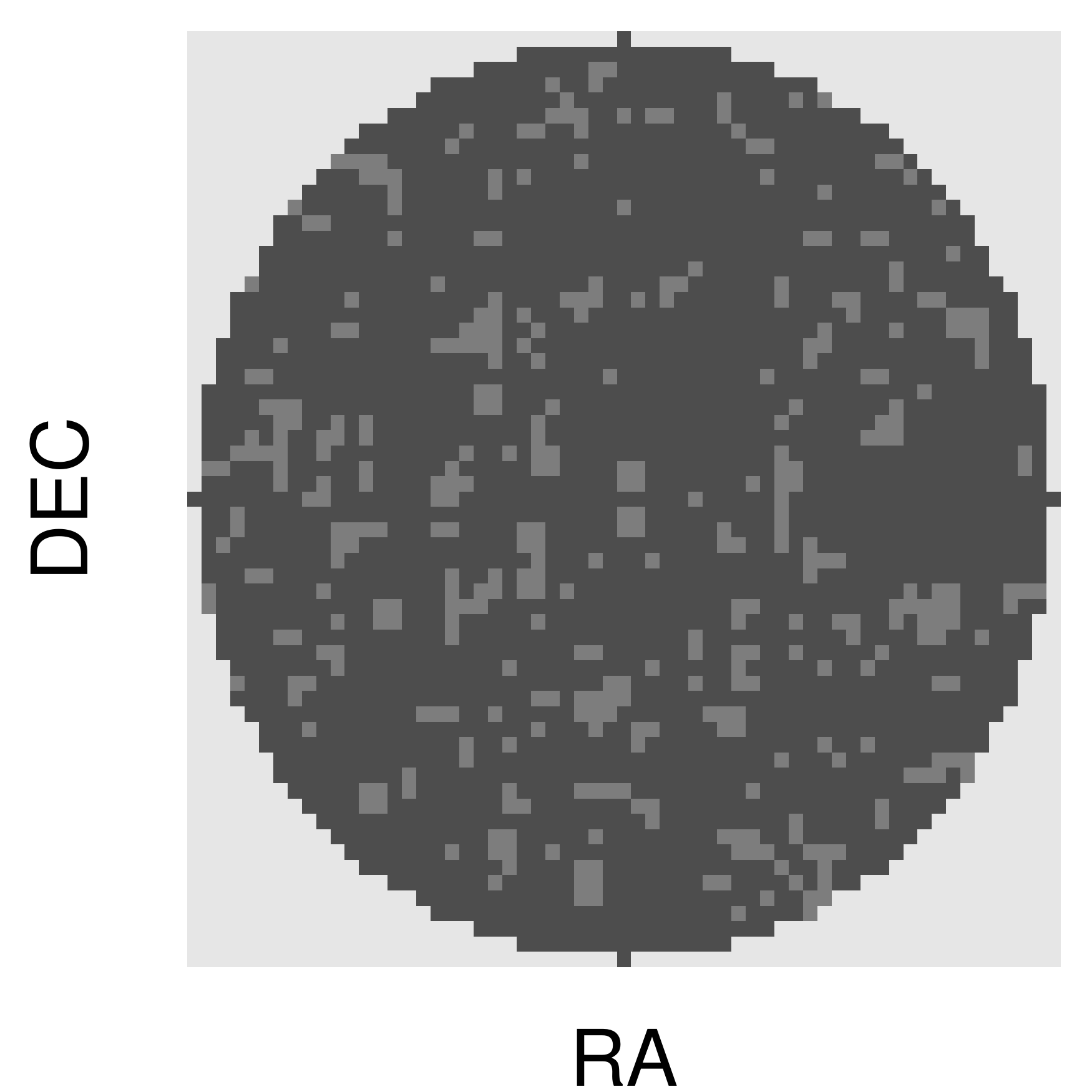}& \includegraphics[width=50mm]{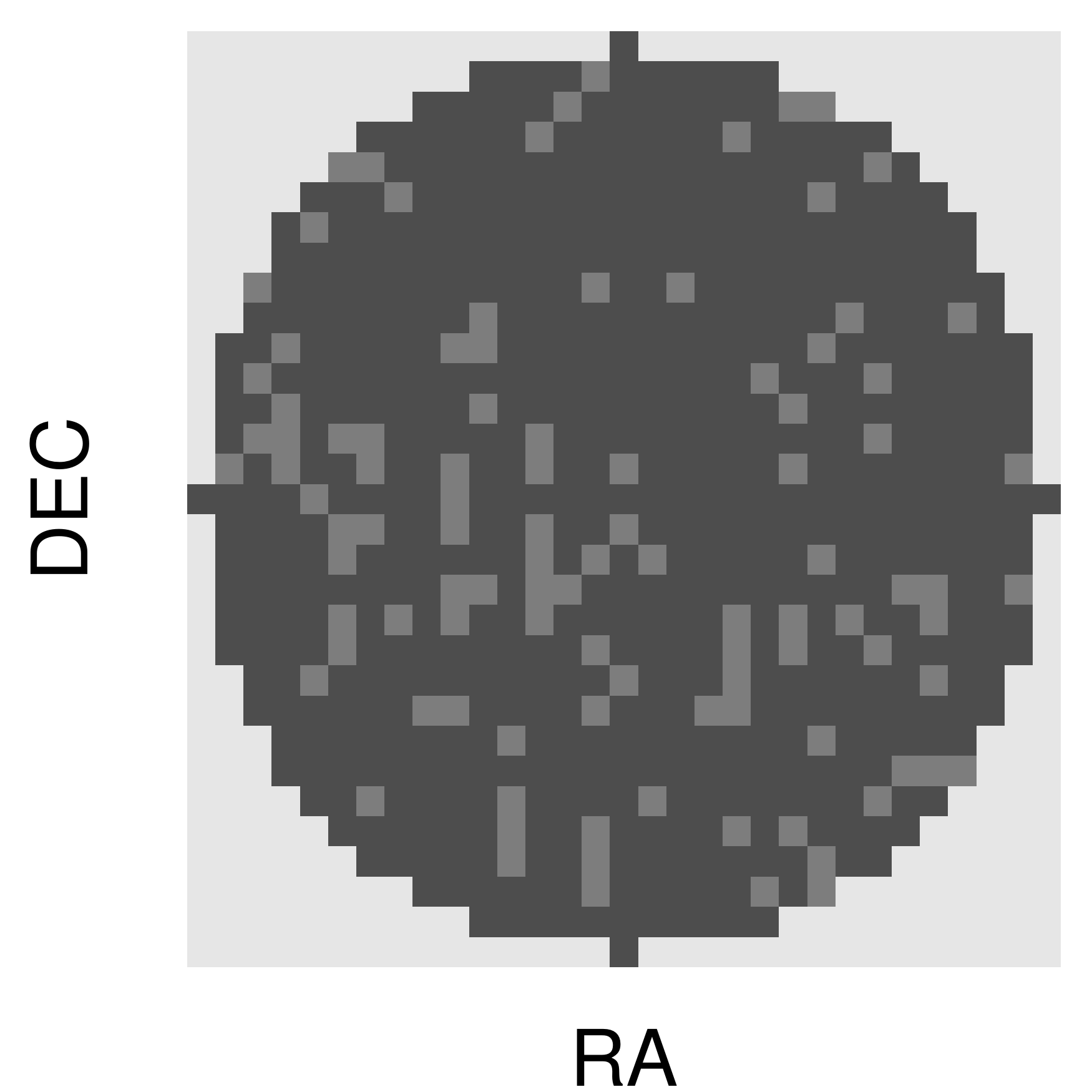} \\
\end{tabular*}
\end{adjustwidth}
\caption{\textcolor{black}{Approximated excursion sets of $W_n(\bm{\theta})$ in Example 1 with respect to $c_k=1$ with $R=14641$, $R=2500$, $R=961$.
The grid points in $\bm{\Theta}_\times$ are chosen at distance 0.5, 1 and 2 in the left, central and right panels, respectively. As $R$ decreases, the excursion set $\mathcal{A}_k$ is poorly approximated by $\mathcal{M}_k$.} }
\label{random_field1}
\end{figure}

\begin{figure}[!h]
\begin{adjustwidth}{0cm}{0cm}
\begin{tabular*}{\textwidth}{@{\extracolsep{\fill}}@{}c@{}c@{}c@{}}
      \includegraphics[width=50mm]{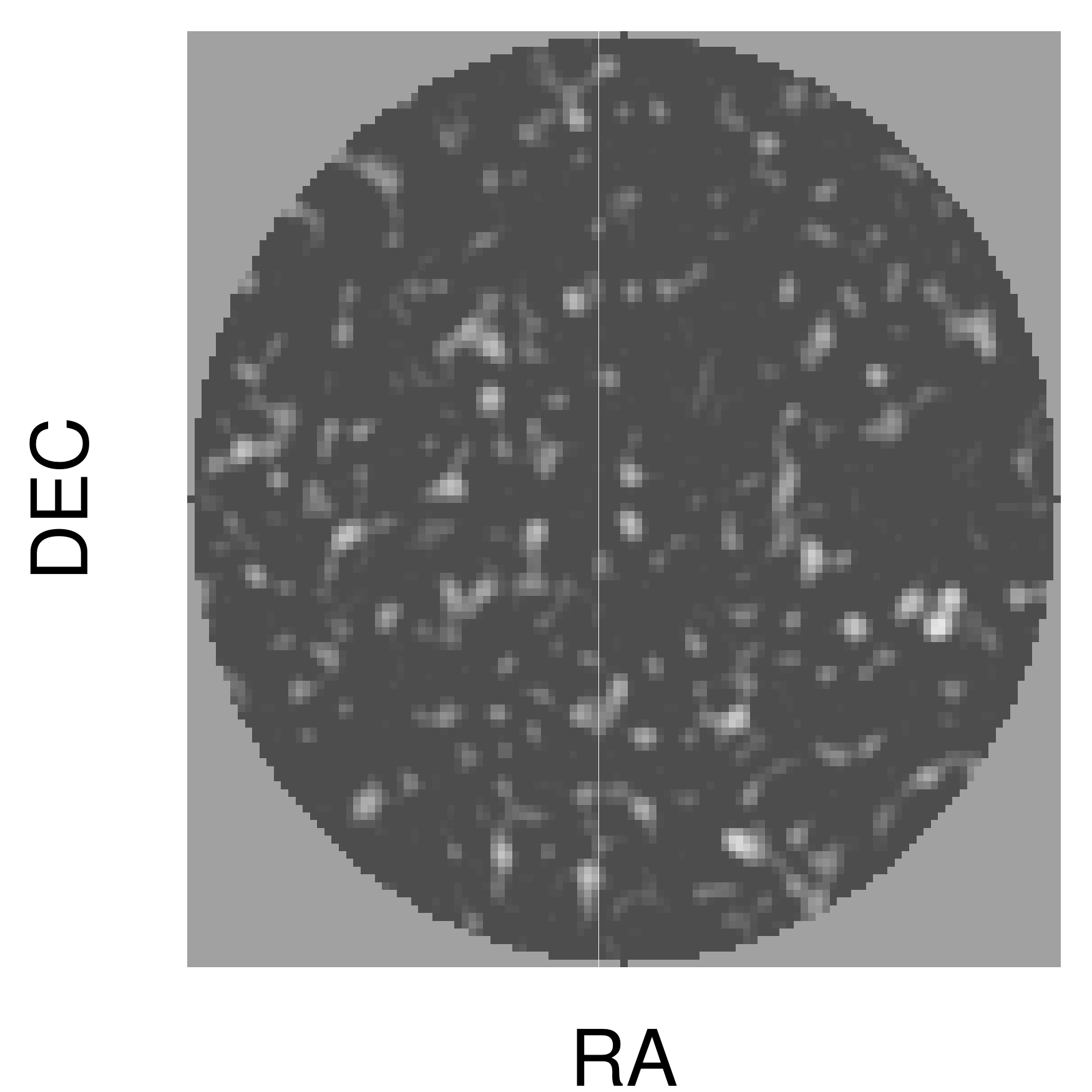} & \includegraphics[width=50mm]{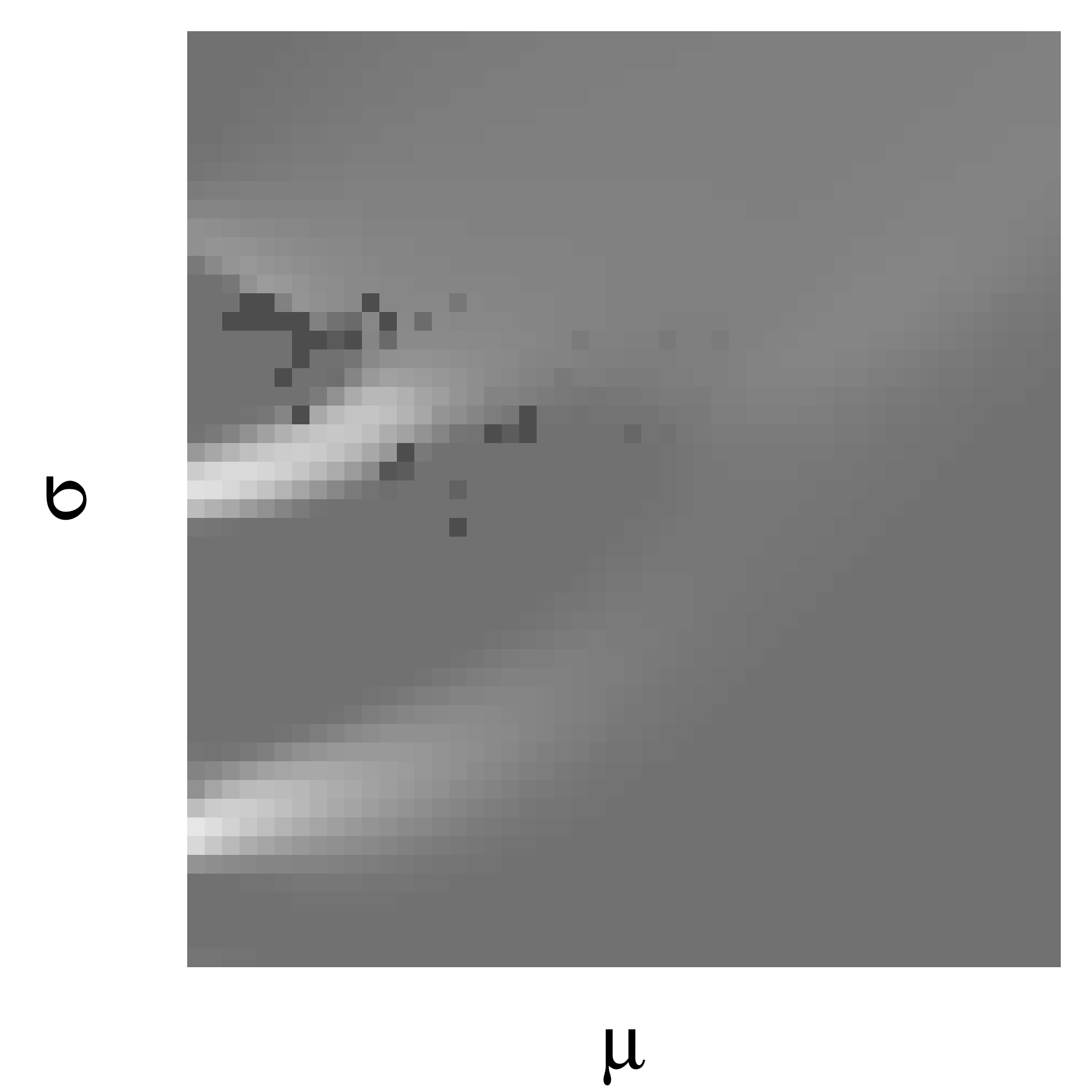}& \includegraphics[width=50mm]{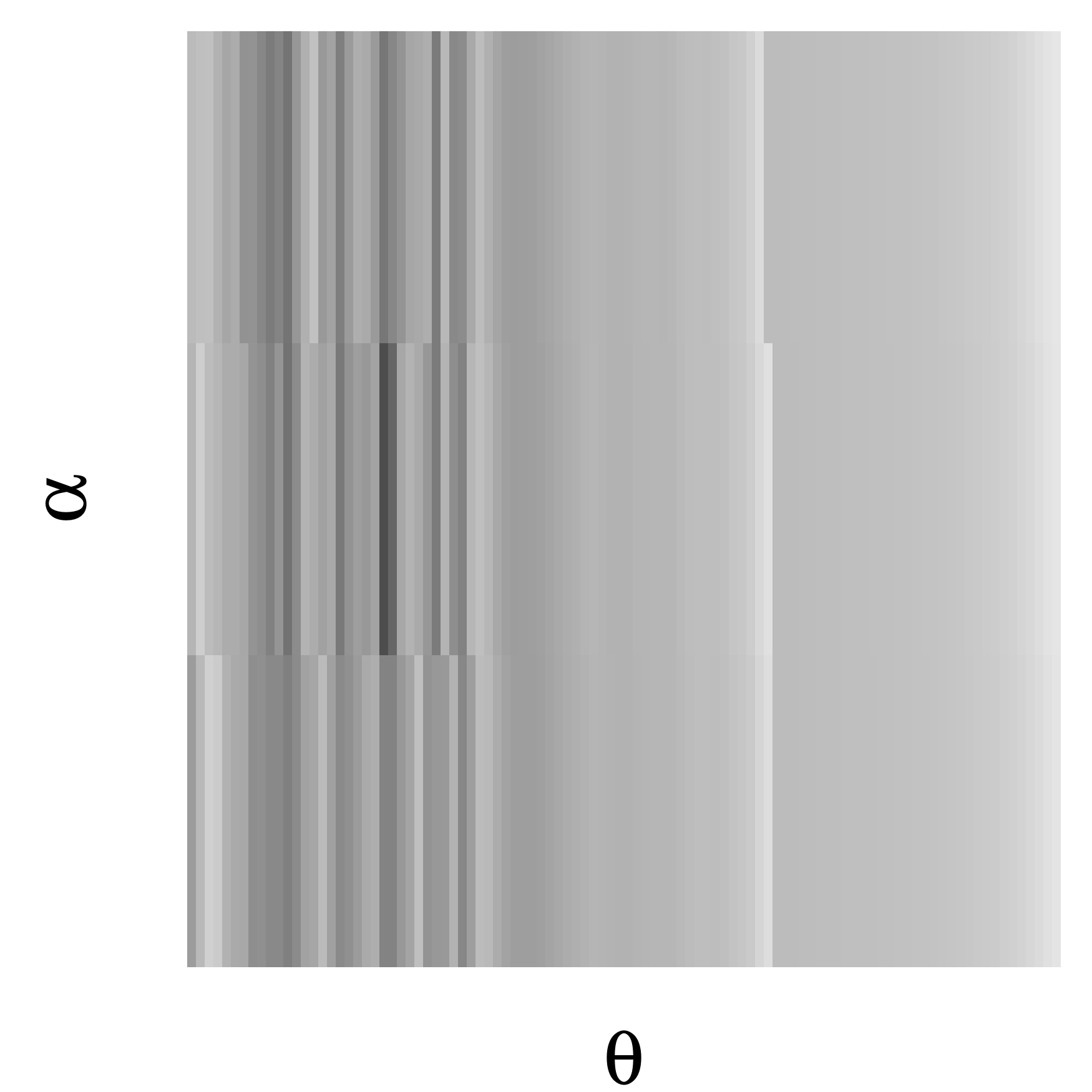} \\
\end{tabular*}
\end{adjustwidth}
\caption{Simulated random fields under $H_0$ for Examples 1, 2 and 3 (left, central, right panels, respectively). }
\label{smoothness}
\end{figure}

\begin{algorithm}[!h]
\label{algo}
\caption{Computing $\phi({\mathcal{A}}_{c_k})$ via graphs} \label{algo}

\begin{enumerate}
\item[ ] \textbf{Input 1:} Constant $c_k$.
\begin{enumerate}
\item[ ] \textbf{Step 1:} For all pairs $(\bm{\theta}_r,\bm{\theta}_s)$ in $\tilde{\mathcal{A}}_{c_k}$  calculate   the distance $d_\varphi(\bm{\theta}_r,                   						\bm{\theta}_s)$ in \eqref{distance};
\item[ ] \textbf{Step 2:} construct the undirected graph $\mathcal{G}^D_{k}=(\tilde{\mathcal{A}}_{c_k}, E^{D}_{k})$ where the edges $E^{D}_{k}$ are allocated according to \eqref{distance2}, with $d=D$;
\item[ ] \textbf{Step 3:} set $j=1$;
\item[ ] \textbf{Step 4:} while  $j<D$:
\begin{enumerate}
		\item[$\qquad$ ] (i) set $d=D-j$;
		\item[ ] (ii) obtain $\mathcal{G}^d_{k}$ from $\mathcal{G}^{d+1}_{k}$ by removing  edges  in $E^{d+1}_{k}$ for which \eqref{distance2} does not hold;
		\item[ ] (iii) count $|C^{d}_{k}|$ in $\mathcal{G}^d_{k}$ via \citet{eppstein};
                    \item[ ] (iv) j=j+1;
\end{enumerate}
		\item[ ]\textbf{Step 5:} calculate $\phi({\mathcal{A}}_{c_k})$ via \eqref{cliques1}.
	\end{enumerate}

\item[ ] \textbf{Output:} Value of  $\phi({\mathcal{A}}_{c_k})$.
	\end{enumerate}
\end{algorithm}
\textcolor{black}{An underlying assumption of our approach is that 
the resolution of $\bm{\Theta}_\times$ is sufficiently high  to guarantee that  $\mathcal{A}_{c_k}$ is  well approximated by $\mathcal{M}_k$. In Example 1 for instance,  choosing a grid of $R=2500$ points (Figure \ref{random_field1}, central panel), leads to a good approximation of $\mathcal{A}_{1}$. Specifically, we    consider as benchmark for the true excursion set $\mathcal{A}_{1}$  a computation of the random field over  a grid of resolution $R=14641$ (left panel of Figure \ref{smoothness}).  Conversely, selecting a grid of size $R=961$ leads to a poor approximation of $\mathcal{A}_{1}$ since several among the connected components disappear (right panel of Figure \ref{smoothness}). Hence, when the size $R$ of the grid is not dictated by the experiment under study, the choice of $R$ should be supported by a sensitivity analysis based on a small simulation of the random field under $H_0$ (e.g., Figure \ref{random_field1}, see also \citet{TOHM}).}

The EC  is  calculated by alternatively adding and subtracting the number of $d$-dimensional hyperrectangles for $d=0,\dots,D$ in $\mathcal{M}_k$ \textcolor{black}{\citep[e.g.,][p.268]{gruber}}. In two dimensions for instance, the EC is obtained by counting the number of vertices, subtracting the number of edges and adding the number of rectangles \citep{worsley95, taylor2008}, e.g., Figure \ref{eulerfig}.

In order to ease  computations in higher dimensions, one   way to count the number of hyperrectangles of arbitrarily large dimension $d$ is summarized in Algorithm \ref{algo} and described below. The goal of Algorithm \ref{algo}  is to construct graphs where the number of  $d$-dimensional complete subgraphs (or cliques, in the second paragraph following) is equal to the number of $d$-dimensional hyperrectangles  in $\mathcal{M}_k$. This can be done as follows.

\begin{figure}[!h]
\begin{adjustwidth}{0cm}{0cm}
\begin{tabular*}{\textwidth}{@{\extracolsep{\fill}}@{}c@{}c@{}c@{}}
      \includegraphics[width=50mm]{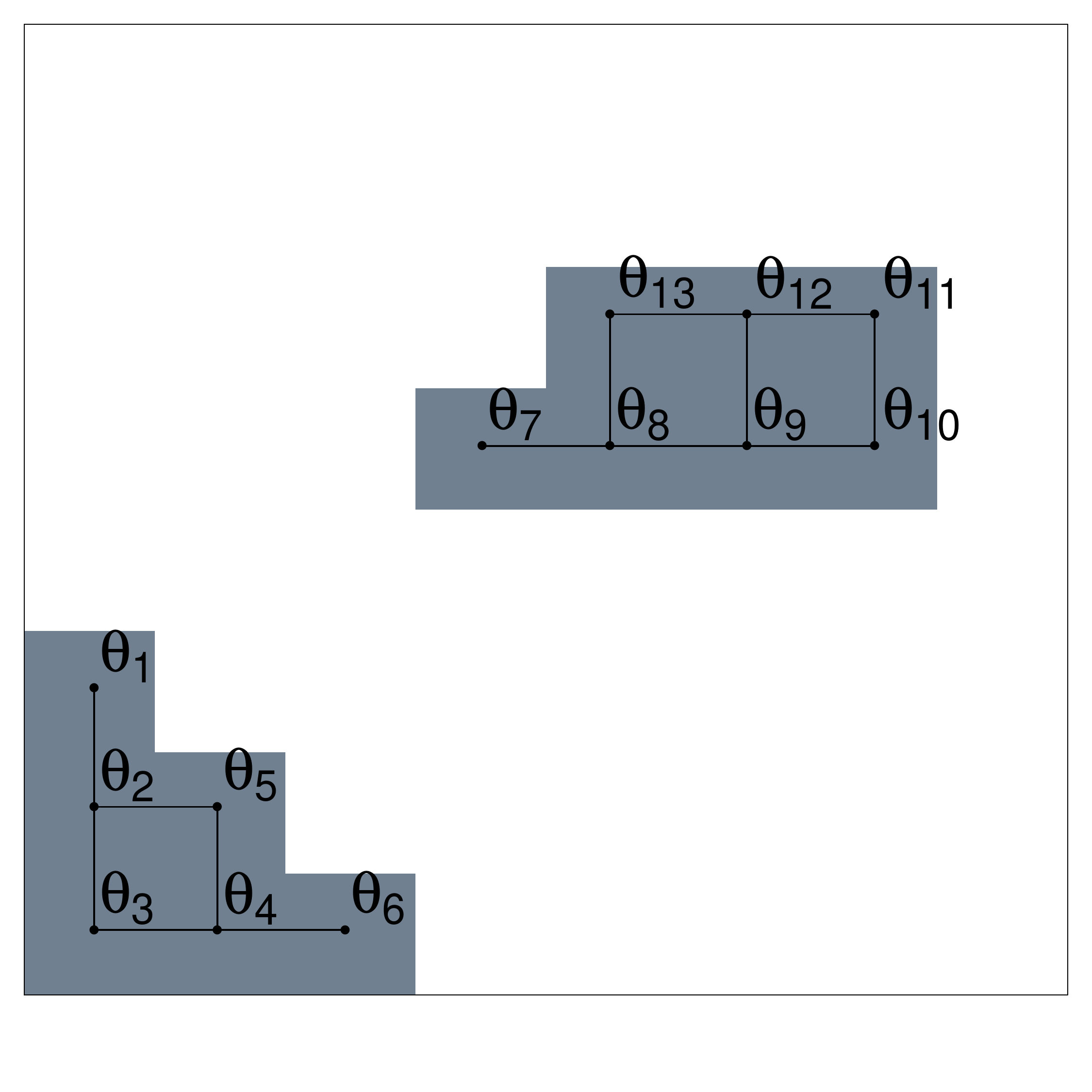} & \includegraphics[width=50mm]{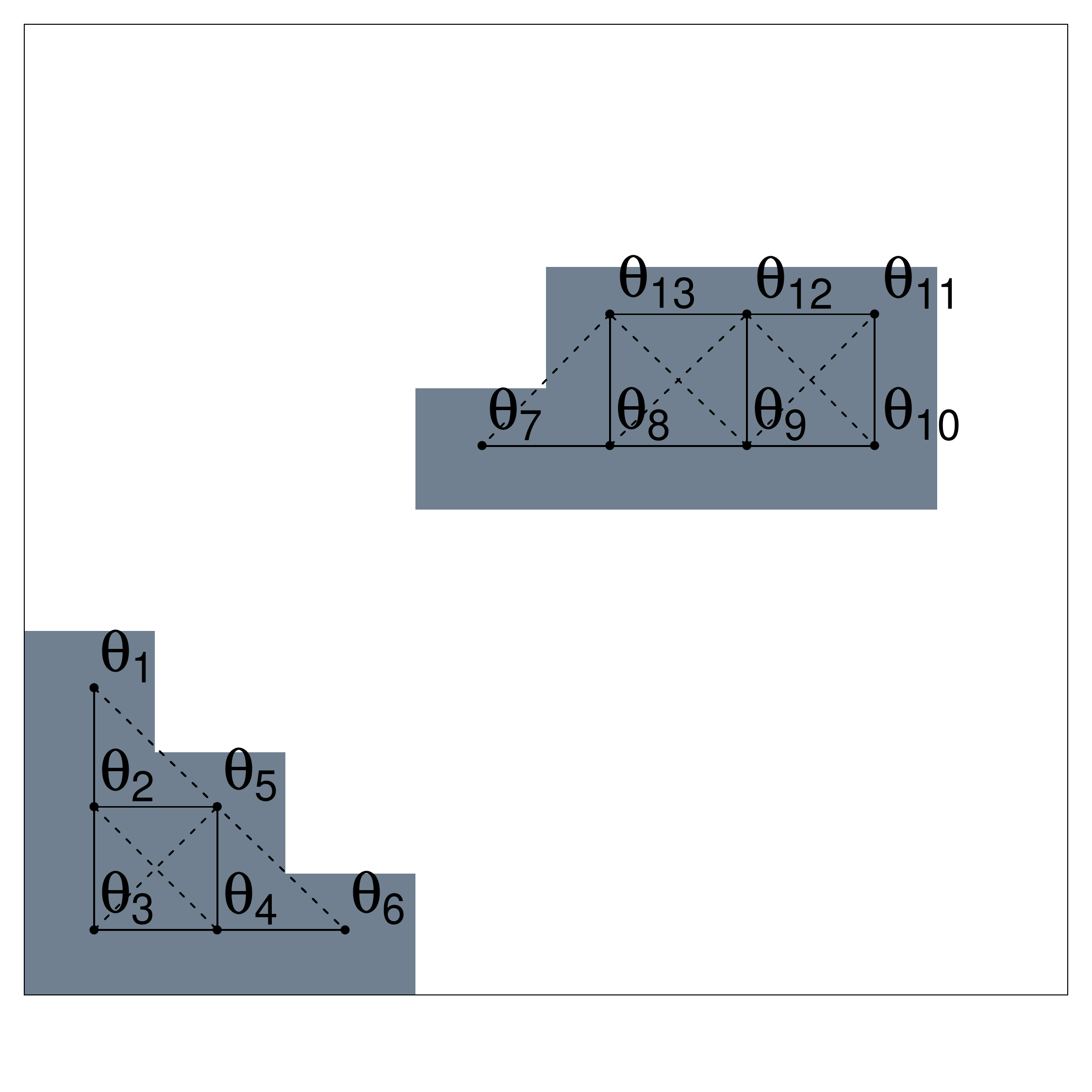}& \includegraphics[width=50mm]{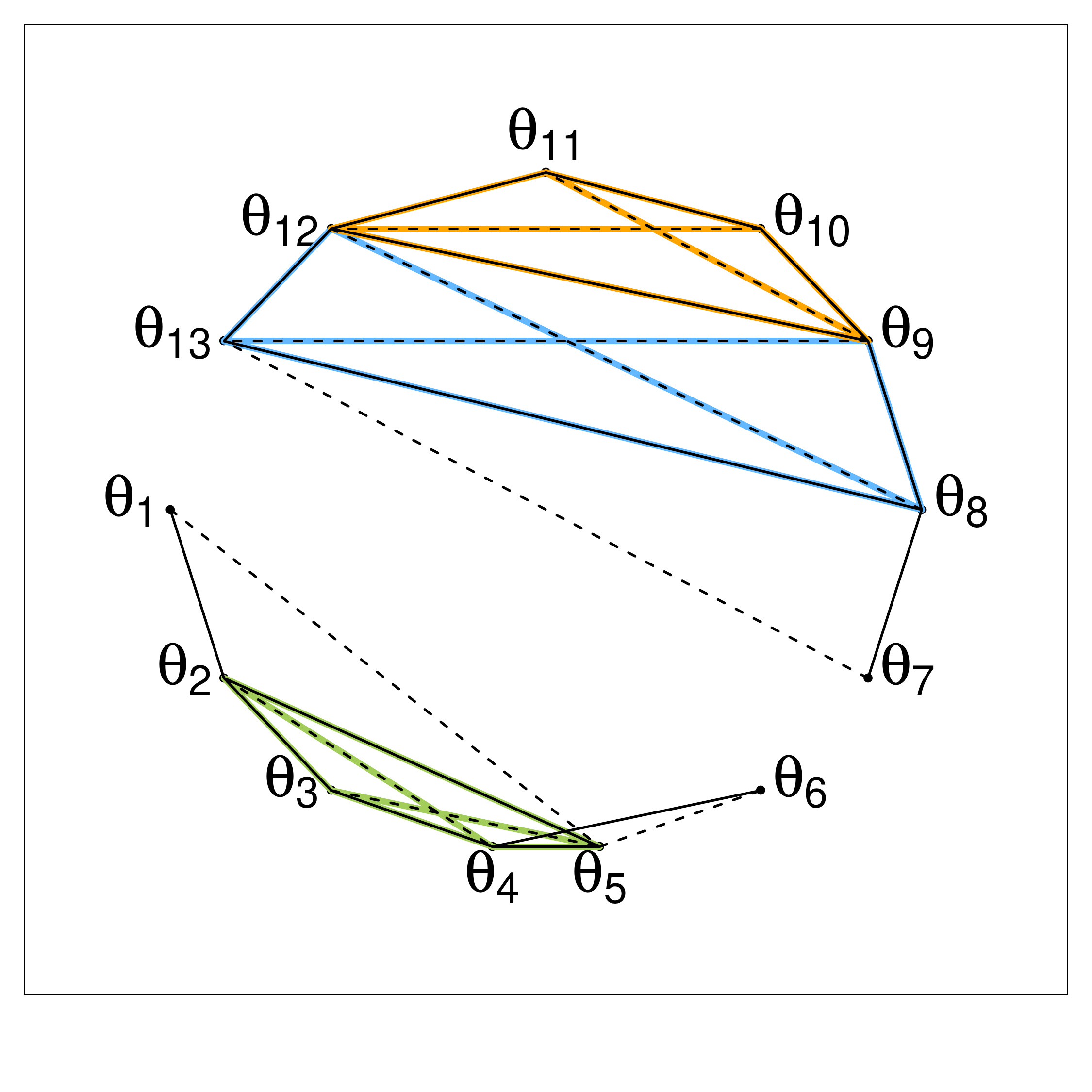} \\
\end{tabular*}
\end{adjustwidth}
\caption{Left panel: quadrilateral mesh ${\mathcal{M}'}_k$ of the excursion set ${\mathcal{A}}_{c_k}$ (gray area), with set of vertices $\tilde{\mathcal{A}}_{c_k}$ (black dots) and edges $E^1_k$ allocated according to \eqref{distance} (black solid segments) of unit length. Central  panel: quadrilateral mesh ${\mathcal{M}'}_k$ and  diagonals of length $\sqrt{2}$ (black dashed segments). Right panel: graph $\mathcal{G}^2_{k}=(\tilde{\mathcal{A}}_{c_k},E^{2}_k)$ in which  the three 4-dimensional cliques in $C^2_k$ are highlighted in orange, blue and green. As expected, each clique in $\mathcal{G}^2_{k}$ corresponds to a square in $\mathcal{M}'_k$.}
\label{graph}
\end{figure}
For each constant $c_k$ in \eqref{system}, and for each dimension  $d=1,\dots,D$,
consider an undirected  unweighted graph, $\mathcal{G}^d_{k}=(\tilde{\mathcal{A}}_{c_k},E^{d}_{k})$, with  vertices  $\tilde{\mathcal{A}}_{c_k}$ and  edges $E^{d}_{k}$ such that two vertices $\bm{\theta}_r$ and $\bm{\theta}_s$ are joined by an edge if and only if
\begin{equation}
\label{distance2}
1\leq\mathtt{d}_{\varphi}(\bm{\theta}_r,\bm{\theta}_s)\leq\sqrt{d},
\end{equation}
where $\sqrt{d}$ corresponds to the length of the longest diagonal of  a $d$-dimensional unit hypercube.

A   graph $\mathcal{G}=(V,E)$ has  a clique of dimension $Q$ if there exists a subset of $Q$ vertices 
in $V$  such that every  pair of distinct vertices of the subset are connected by an edge. We denote the set of all $2^d$-dimensional cliques in  $\mathcal{G}^d_{k}$ by  $C^d_{k}$. 
The distance between points in $\tilde{\mathcal{A}}_{c_k}$ does not affect the enumeration of the hyperrectangles in $\mathcal{M}_k$. Specifically, since  the allocation of the edges $E^1_{k}$  only depends on the indexes
$\varphi_d{(r)}$ of the $\theta_{rd}$ within $\tilde{\Theta}_{d}$, for $d=1,\dots,D$,  the number of $d$-dimensional hyperrectangles  in  $\mathcal{M}_{k}$ is equal to the number of $d$-dimensional unit hypercubes in a ``unit'' mesh, denoted by $\mathcal{M}'_{k}$,  with   vertices $\tilde{\mathcal{A}}_{c_k}$ and edges ${E}^1_{k}$  of unit  length.

It follows that the  $2^d$ vertices of  each clique in $C^d_{k}$ is a subset of  points  in $\tilde{\mathcal{A}}_{c_k}$ which are at least one unit, and at most $\sqrt{d}$, apart one another.  By construction, this implies that each  clique in $C^d_{k}$ corresponds to a unit $d$-dimensional hypercube in  
$\mathcal{M}'_{k}$, which in turn corresponds to a $d$-dimensional hyperrectangle in $\mathcal{M}_{k}$. For illustrative purposes, in Figure \ref{graph} we give an example in two dimensions, where for simplicity the points $\bm{\theta}_r$ are equally spaced over  unit intervals in each $\tilde{\Theta}_{d}$, $d=1,2$, and thus   $\mathcal{M}_{k}=\mathcal{M}'_{k}$\footnote{ Notice that the main difference between the  mesh $\mathcal{M}_{k}$ (or $\mathcal{M}'_{k}$) and the graph  $\mathcal{G}^D_{k}$ is that the former depends on  the position of  its vertices  in  $\bm{\Theta}$ and their distance; whereas the latter only accounts for their connectivity. }.

Therefore, in general terms, we can compute $\phi({\mathcal{A}}_{c_k})$ as
\begin{align}
\label{cliques1}
\phi({\mathcal{A}}_{c_k})&=\sum_{d=0}^{D}(-1)^d|C^d_{k}|\\
\vspace{-0.5cm}
\label{cliques2}
&=|\tilde{\mathcal{A}}_{c_k}|-|E_{k}|+\sum_{d=2}^{D}(-1)^d|C^d_{k}|
\end{align}
where $|\cdot|$ is the cardinality of the set considered. Equation \eqref{cliques2} follows from \eqref{cliques1} since by construction $\mathcal{G}^0_k=\tilde{\mathcal{A}}_{c_k}$, $\mathcal{G}^1_k$ is the unweighted graph with the same vertices and edges of $\mathcal{M}_{k}$ and $\mathcal{M}'_{k}$; thus $|C^0_{k}|=|\tilde{\mathcal{A}}_{c_k}|=\sum_{r=1}^R\mathbbm{1}_{\{w(\bm{\theta}_r)>c_k\}}$ and $|C^1_{k}|=|E^1_{k}|=\sum_{r=1}^R\sum_{s=1}^R\mathbbm{1}_{\{\mathtt{d}_\varphi(\bm{\theta}_r,\bm{\theta}_s)=1\}} $.

Naively, computing $|C^d_{k}|$ by sequentially considering each subset of $\tilde{\mathcal{A}}_{c_k}$ of size $2^d$  requires a complexity $O(|\tilde{\mathcal{A}}_{c_k}|^{2^D} 4^D)$ to evaluate \eqref{cliques1}, a massive computation load unless $D$ is quite small.
The advantage of converting the hyperrectangles enumeration problem into a clique-finding problem  is that several efficient algorithms exists to address this challenge in near-optimal time \citep[e.g.,][]{bron, johnston,eppstein}. In our implementations in Section \ref{sim},
we use the algorithm proposed by \citet{eppstein}, and   implemented in the \verb R  function \verb cliques    in the \verb igraph   package \citep{igraph}.  Specifically, \citet{eppstein} propose a variation of the Bron-Kebosch algorithm \citep{bron} for sparse graphs where the running time is of $O(h|\tilde{\mathcal{A}}_{c_k}|^{\frac{h}{3}})$, with $h=2^D-1$. This is particularly convenient in our context where the constants $c_k$ can be chosen arbitrarily to reduce both the size of the graph and its sparsity. 
Hence in Algorithm \ref{algo} we recommend a top-down approach where $\mathcal{G}^D_{k}$ is constructed first, and
the constants $c_k$ can be adequately adjusted between Step 2 and Step 3   in order to increase sparsity in  $\mathcal{G}^D_{k}$.
The graphs $\mathcal{G}^d_{k}$, for $d=0,\dots,D-1$, are obtained subsequently by removing  edges for which \eqref{distance2} is not satisfied as $d$ decreases. An additional advantage of this approach is that $\mathcal{G}^D_{k}$  provides a simple two-dimensional representation of the $D$-dimensional excursion sets ${\mathcal{A}}_{c_k}$.

Finally, since in practice the $E[\phi(\mathcal{A}_{c_k})]$ are unknown, we can estimate them via Monte Carlo simulation.
Specifically, for  \textcolor{black}{arbitrary choices of $c_k$}, Monte Carlo estimates of $E[\phi(\mathcal{A}_{c_k})]$, namely $\widehat{E[\phi({\mathcal{A}}_{c_k})]}$, can  be obtained via Algorithm \ref{algo} with a small set of Monte Carlo replicates of $\{W(\theta_r)\}$ and averaging over the  values $\phi({\mathcal{A}}_{c_k})$ obtained at each replicate. The reader is referred to Section \ref{sim} for a discussion on the accuracy of $\widehat{E[\phi({\mathcal{A}}_{c_k})]}$.
\vspace{-1cm}
\begin{figure}[H]
\begin{adjustwidth}{0cm}{0cm}
\begin{tabular*}{\textwidth}{@{\extracolsep{\fill}}@{}c@{}c@{}c}
   \multicolumn{2}{c}{  \includegraphics[width=90mm]{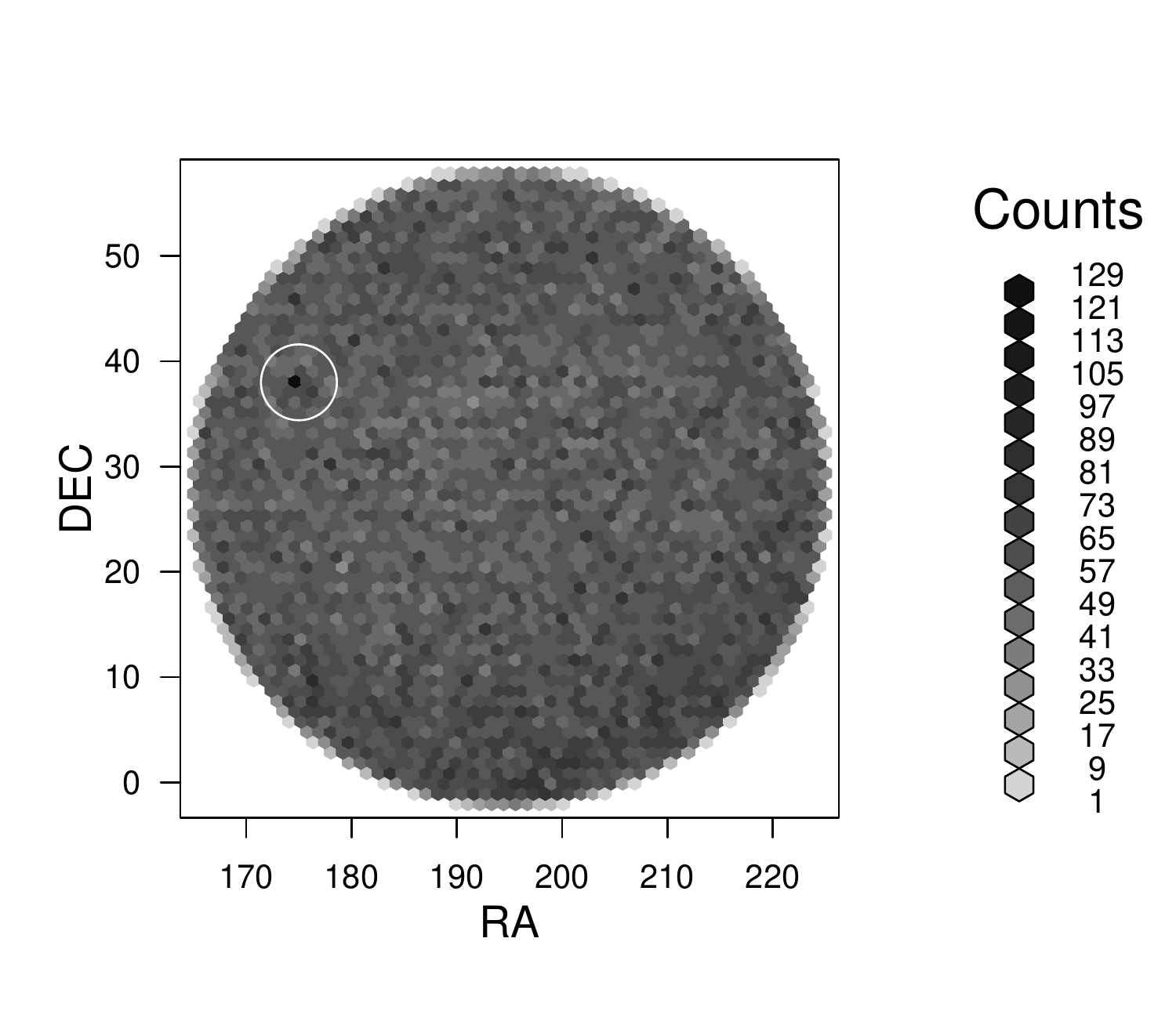}} \\
 \includegraphics[width=65mm]{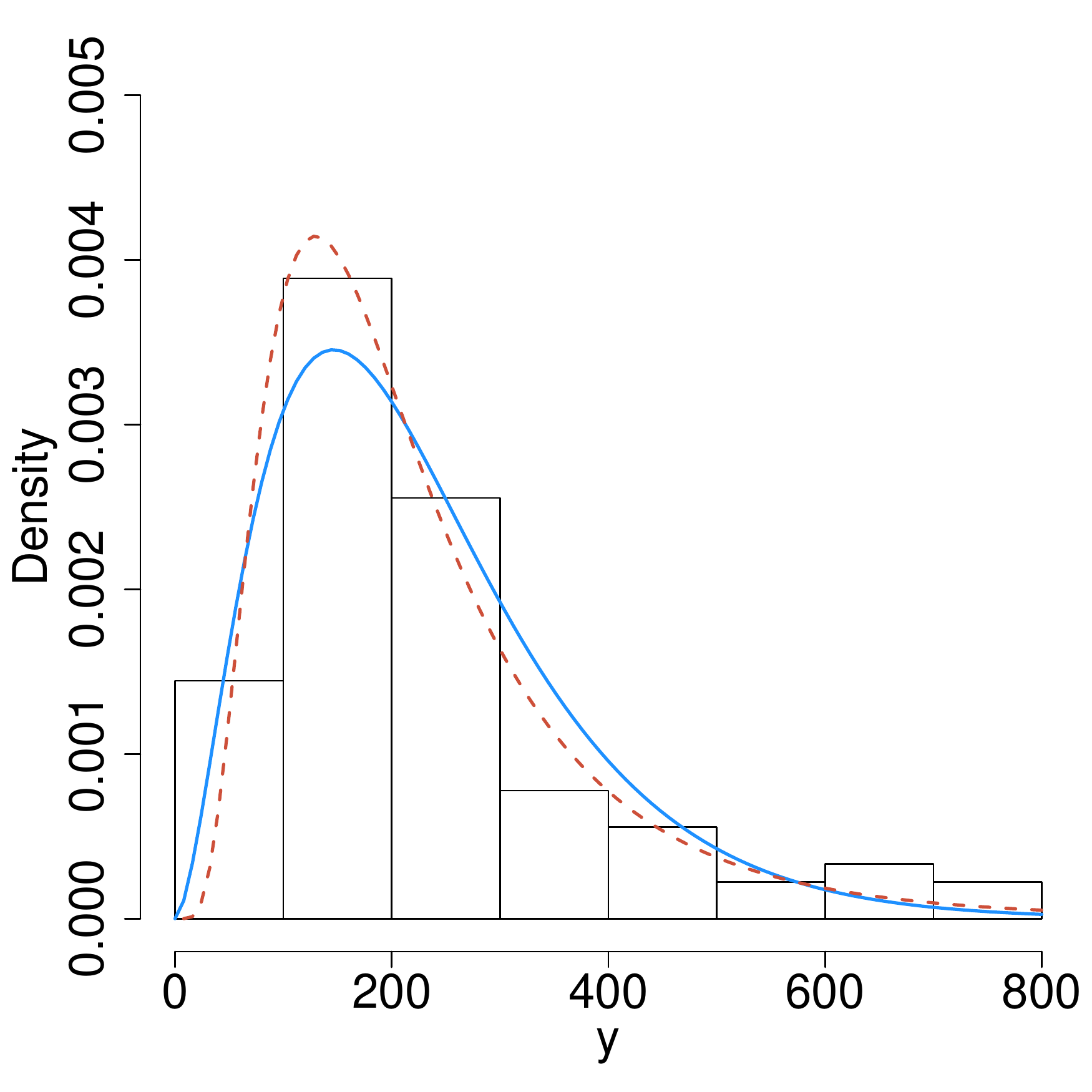}& \includegraphics[width=65mm]{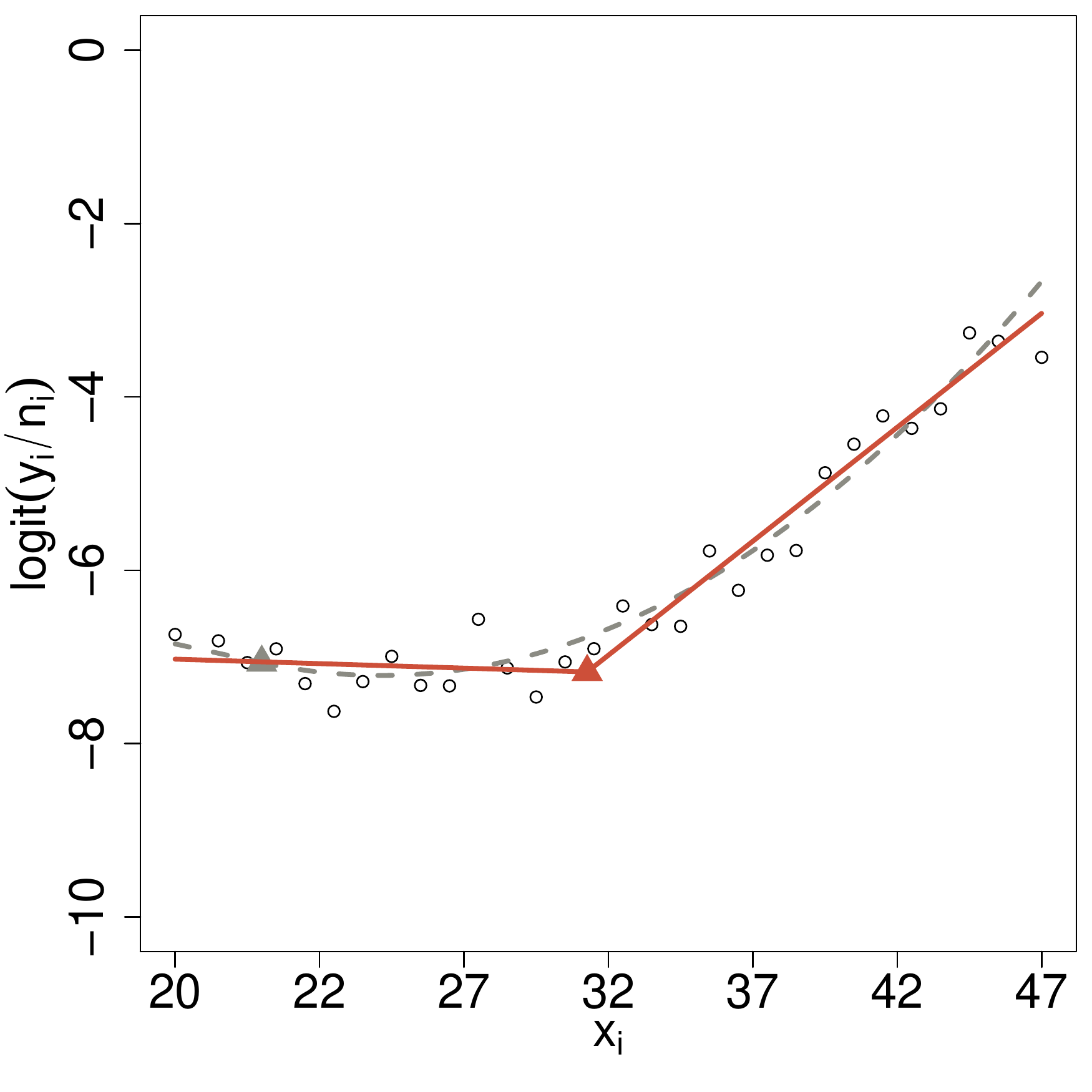} \\
\end{tabular*}
\end{adjustwidth}
\caption{Top panel: 2D histogram of the Fermi-LAT realistic data simulation for Example 1. The white circle indicates the  location at which the LRT-process achieves is maximum, i.e., ${\bm{\theta}}=(175,38)$ with estimated intensity $\hat{\eta}=0.001$. Bottom left panel: histogram of   maize seeds strength in Example 2. The null model in \eqref{test1} (blue solid curve) is fitted as a gamma distribution with $(\hat{\tau},\hat{\gamma})=(2.762,83.007)$.
The null model in \eqref{eta1} (red dashed line) is fitted as a log-normal distribution with $(\hat{\mu},\hat{\sigma})=(5.243,0.614)$. 
Bottom right panel: Down syndrome data, the model in \eqref{logistic} selected by THOM is a break point logistic regression with linear trend (red   solid lines) i.e., $\alpha=1$ and ${\theta}=31.265$   (red triangle). For comparison,  a  break point logistic regression with change of trend from linear to quadratic (gray dashed line) is also fitted while fixing $\alpha=2$. In this case the breakpoint occurs at ${\theta}=20$ (gray triangle).}
\label{real_plots}
\end{figure}
Consequently, we can  estimate the right hand sides of   \eqref{technical_bound3} with
\begin{equation}
\label{replacement1}
{\mathcal{L}}_0(\bm{\Theta})P(W(\bm{\theta}_1)>c)+\sum^{D}_{j=1}\widehat{\mathcal{L}^*_d(\bm{\Theta})}\rho_d(c)\
\end{equation}
where $\widehat{\mathcal{L}^*_d(\bm{\Theta})}$ are the solution of the system of equation in \eqref{system} with $E[\phi({\mathcal{A}}_{c_k})]$ in the left hand sides of each equation  replaced by their Monte Carlo estimates $\widehat{E[\phi({\mathcal{A}}_{c_k})]}$.

\section{Numerical results}
\label{sim}  
\subsection{Case studies: description}
\label{casestudies}  
In this section we apply TOHM to the three examples introduced in Sections \ref{intro} and \ref{main}, i.e., a dark matter signal search, a non-nested model comparison and a logistic regression with a break point and change of trend.

\begin{figure}[!h]
\begin{adjustwidth}{0cm}{0cm}
\begin{tabular*}{\textwidth}{@{\extracolsep{\fill}}@{}c@{}c@{}c@{}}
      \includegraphics[width=50mm]{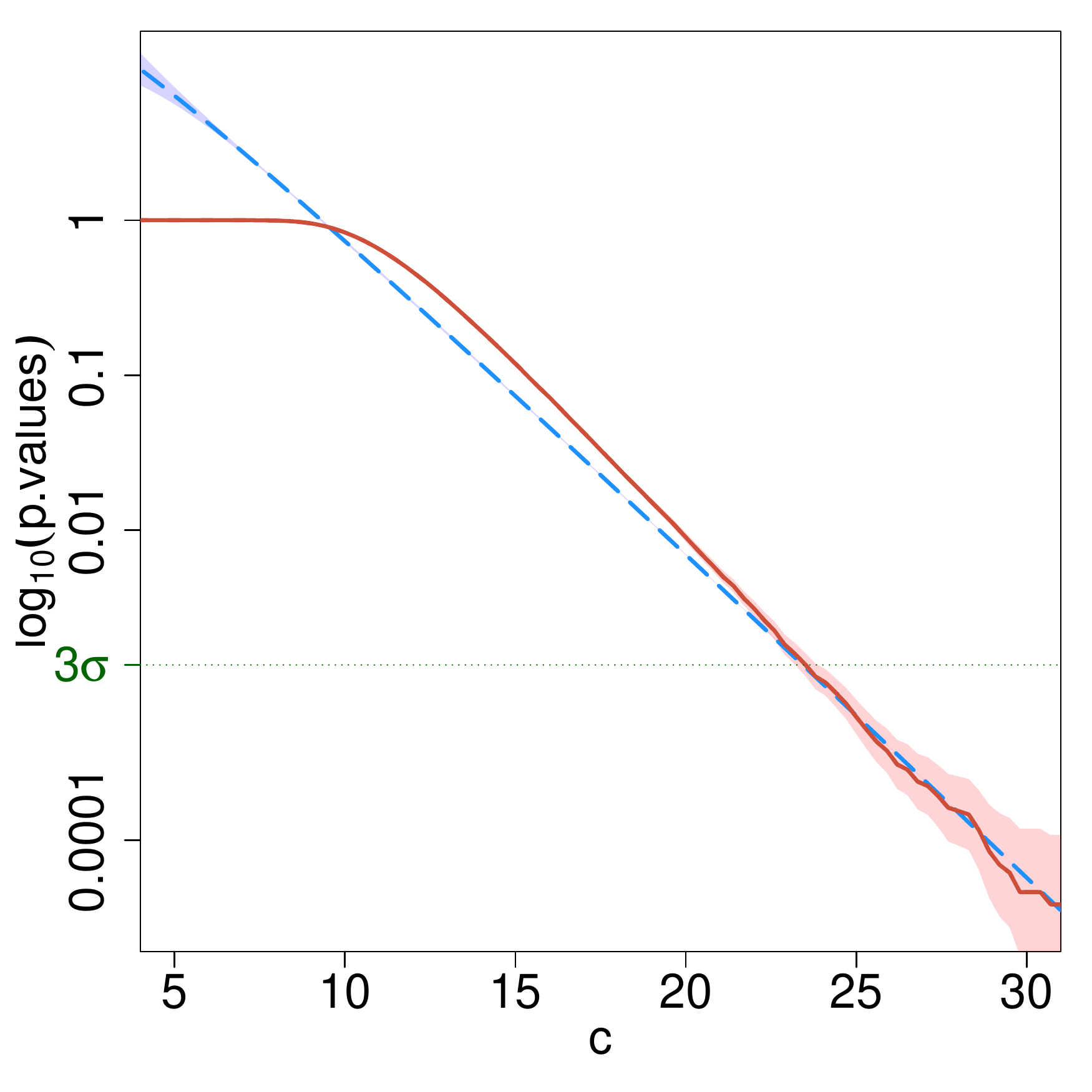} & \includegraphics[width=50mm]{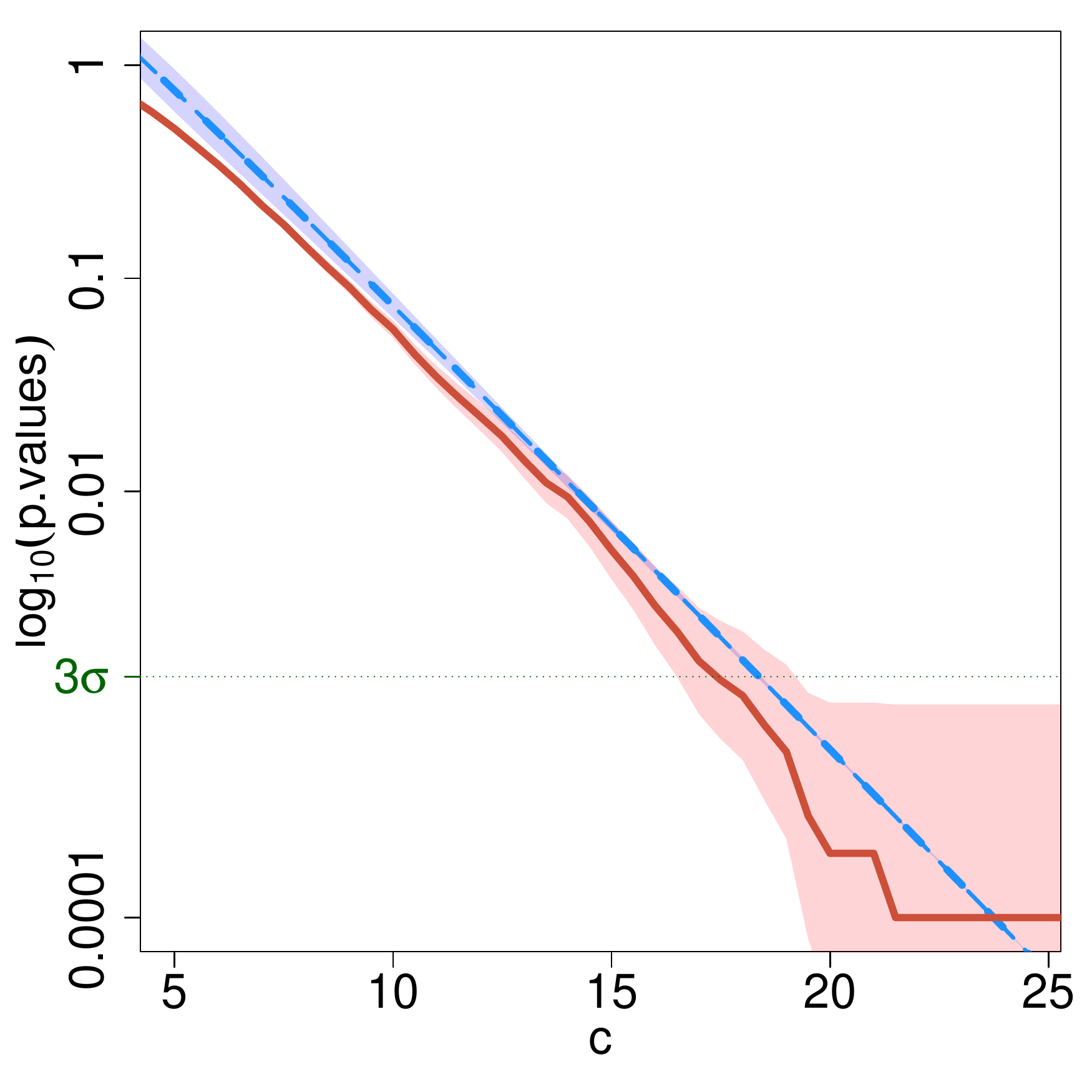}& \includegraphics[width=50mm]{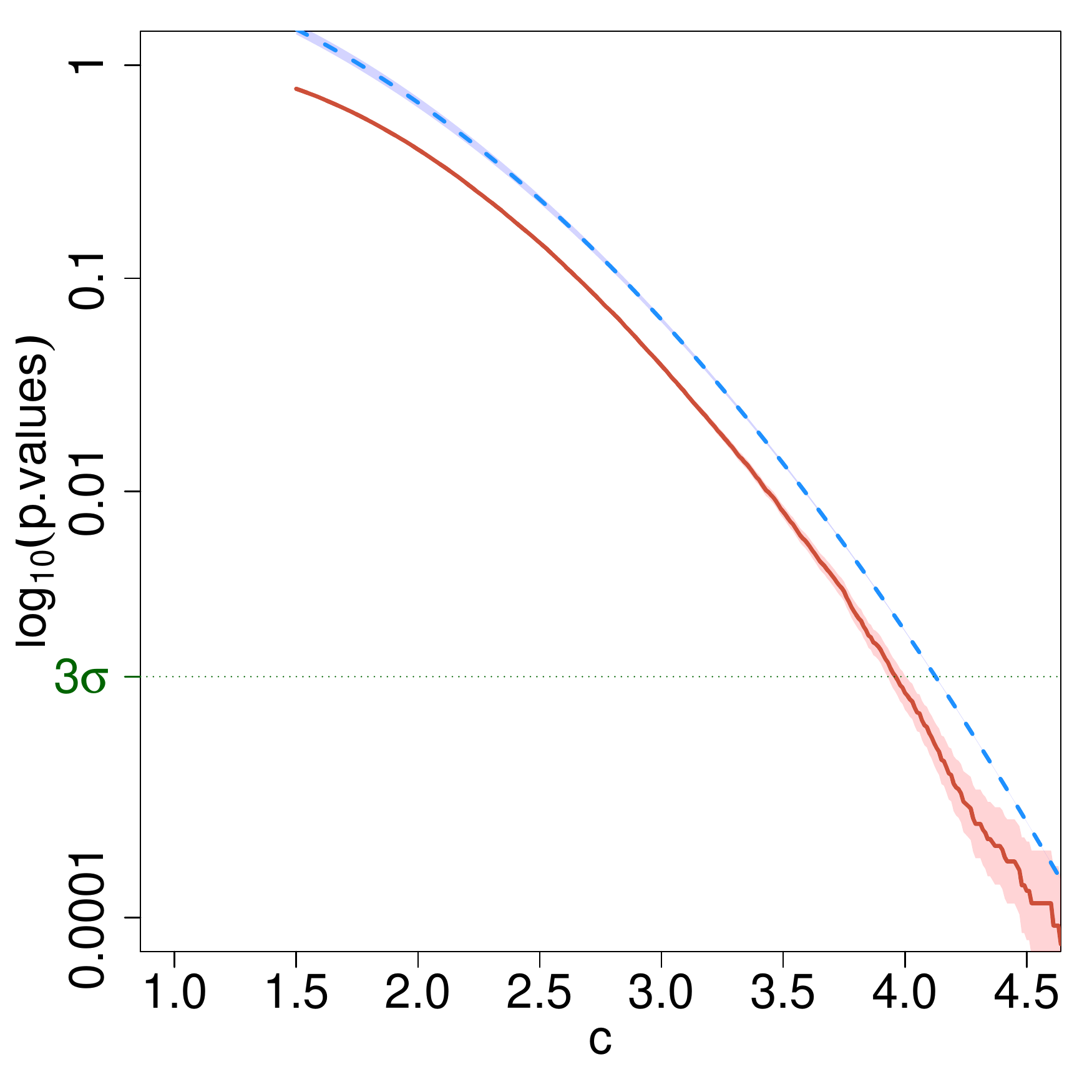} \\
\end{tabular*}
\end{adjustwidth}
\caption{Estimated   approximations in \eqref{replacement1} (blue dashed  line), Monte Carlo estimates of $P(\sup_{\bm{\theta} \in \bm{\Theta}}\{W(\bm{\theta})\}>c)$ (red solid line) in $log_{10}$-scale,  and   Monte Carlo errors (pink areas) for increasing values of the threshold $c$, for Example 1 (left panel), Example 2 (central panel) and Example 3 (right panel). Monte Carlo errors associated with 
$\widehat{E[\phi({\mathcal{A}}_{c_k})]}$  in \eqref{replacement1} are plotted as gray areas.  }
\label{assess}
\end{figure}
In Example 1, we consider a realistic simulation of the Fermi Large Area Telescope (LAT) 
obtained with the \emph{gtobssim} package\footnote{\url{http://fermi.gsfc.nasa.gov/ssc/data/analysis/software}}. 
This analysis represents a simplified example in the context of searches for $\gamma$-ray lines in galaxy clusters \citep{refB2,refB1,refB3}.
Our goals are  (i) to assess the presence of a
 photon emission due to a dark matter source in addition to cosmic background photons, and 
(ii) to identify the location at which maximum evidence in favor of the  suspected source is achieved. \textcolor{black}{ The cosmic background is uniformly distributed over  the search region $\bm{\Theta}$ which in this case corresponds to
a disc  in the sky of $30^\circ$ radius and centered at  ($195$ RA,$28$ DEC), where RA and DEC are the coordinates in the sky}, and thus in \eqref{unif_gauss}  $x\in[165,195]$, $y\in[28-\sqrt{30^2-(x-195)^2}, 28+\sqrt{30^2-(x-195)^2}]$. In our simulation the dark matter source emission  is located at $(\theta_1,\theta_2)=(174.952, 37.986)$ and realistic representations of the systematic errors, as well as the calibration of the detector, are included.
This set up led to $139,821$ background events and $144$ dark matter events; these data are available in the Supplementary Materials.

In Example 2, we apply TOHM to the \emph{Compressive strength and strain of maize seeds dataset} available in  the R package \verb goft   \citep{goft}. The dataset records the compression strength in Newtons of $90$ seeds and the goal is to choose between a gamma and a  log-normal distribution for the data. In order to ease our computations,  in \eqref{nonnest} we let   $y\in (0,1000]$ and  $(\mu,\sigma^2)=[1,10]\times[0.2,5]$.

Finally, in Example 3 we consider the \emph{Down Syndrome dataset} available in the R package \verb segmented    \citep{segmented}. The dataset 
records whether babies born to 354,880 women are affected by Down Syndrome. Our goal is to use TOHM to assess the presence of a break point when regressing the logit of the probability $\pi_i$ that a woman of age $x_i$ delivers a baby with down syndrome, where $x_i\in[17,47]$,  and we let $\theta\in[20,44]$. In contrast to the analysis in \citet{TOHM} we  allow  a change of trend after the break point. Specifically, we allow for a quadratic trend, a change of the linear trend or a break due to a change of the intercept, i.e., $\alpha\in\{0,1,2\}$. 
The data for Examples 1-3 are plotted in Figure  \ref{real_plots}.


\subsection{Goodness of the approximations}
\label{goodness}
Our first task is to assess the accuracy of the approximation of  $P(\sup_{\bm{\theta} \in \bm{\Theta}}\{W(\bm{\theta})\}>c)$ in \eqref{replacement1}, as $c\rightarrow\infty$. 

\textcolor{black}{
In Examples 1 and 2, $\{W(\bm{\theta})\}$ is a function of a zero mean and unit variance  Gaussian random field $\{Z(\bm{\theta})\}$, whereas in Example 3, $\{W_n(\bm{\theta})\}$ is asymptotically distributed as $\{Z(\bm{\theta})\}$  (see Propositions \ref{prop1} and \ref{prop2}).  As discussed in footnote 2, among the conditions of  \citet[][p.347]{taylor2003} which guarantee the validity of  \eqref{lipscitz}, $\{Z(\bm{\theta})\}$ must have almost surely  continuous partial derivatives up to the second order, and the two-tensor field induced by $\{Z(\bm{\theta})\}$ must be not degenerate over $\bm{\Theta}$. These assumptions guarantee smoothness of $\{W(\bm{\theta})\}$  but unfortunately, they are often difficult to verify directly  as they  require knowledge of both the joint distribution of the random fields and their derivatives. In this article, we  limit our  assessment of  the smoothness of $\{W(\bm{\theta})\}$   to a small Monte Carlo simulation, and in Figure \ref{smoothness} we report the results of one of the Monte Carlo replicates obtained for each of the examples considered. }

\textcolor{black}{Both Examples 1 and 2 exhibit smooth random fields (left and central panel) under $H_0$. This is not surprising since the  covariance function of the underlying Gaussian process $\{Z(\bm{\theta})\}$  can be specified as a function of the respective Fisher information matrix \citep[][p.16]{ghosh}, and which can be shown to be twice differentiable (see page 4 of the Supplementary Material).  Conversely,  the signed-root LRT random field in Example 3 appears to be particularly irregular.
This is due to the fact that $\{W_n(\bm{\theta})\}$ in \eqref{gLRT} is continuous, but not continuously differentiable with respect to the parameter $\theta$,  which indexes the location of the break-point. Keeping $\alpha$ fixed, the EC heuristic is not affected by this lack of regularity if the number  of jumps in the derivative with respect to  $\theta$   is finite  \citep[see][]{davies87}; however, when considering $\bm{\theta}=(\theta,\alpha)$,  the number of jumps easily diverges if the parameter space of $\alpha$ is continuous. In Example 3, this effect is mitigated by choosing  $\alpha\in\{0,1,2\}$, but this introduces a different  source of non-regularity since $\{W_n(\bm{\theta})\}$ is no longer continuous with respect to $\alpha$. Therefore, it is particularly interesting to assess if \eqref{replacement1} can still be used as a reliable approximation for $P(\sup_{\bm{\theta} \in \bm{\Theta}}\{W(\bm{\theta})\}>c)$ despite this lack of regularity. }

\begin{table}[!h]
\fontsize{9}{11}\selectfont{
 \centering
\begin{tabular}{c|c|c|c}
\noalign{\global\arrayrulewidth0.05cm}
 \hline
 \noalign{\global\arrayrulewidth0.05pt}
 &  &   &    \\
\bf{Example}& \bf{Selected}& \bf{P-value} & \bf{\textcolor{black}{Monte Carlo Error} } \\
&${\bm{\theta}}$ &\bf{(significance)}  & \bf{(significance interval)} \\
 &  &   &    \\
   \noalign{\global\arrayrulewidth0.05cm}
 \hline
 \noalign{\global\arrayrulewidth0.08pt}
 &  &   &    \\
Example 1& $(\theta_1,\theta_2)$ & $1.092\cdot10^{-26}$& $9.272\cdot10^{-28}$\\
& $(175, 38)$&$(10.629\sigma)$& $[10.621\sigma,10.637\sigma]$\\
 &  &   &    \\
\hline
 \noalign{\global\arrayrulewidth0.08pt}
 &  &   &    \\
Example 2 &  $(\mu,\sigma)$ & $0.034$& $0.012$\\
$H_0:\eta=0\quad\text{vs}\quad H_0:\eta>0$&  $(5.041, 0.5)$ &$(1.801\sigma)$&$[1.663\sigma,1.988\sigma]$\\
 &  &   &    \\
\hline
 \noalign{\global\arrayrulewidth0.08pt}
 &  &   &    \\
Example 2& $(\gamma,\tau)$ & \textcolor{black}{$0.596$} & \textcolor{black}{$0.093$} \\
$H_0:\eta=1\quad\text{vs}\quad H_0:\eta<1$&\textcolor{black}{$(6.510, 70)$} &$(0.00\sigma)$&\\
 &  &   &    \\
\hline
 \noalign{\global\arrayrulewidth0.08pt}
 &  &   &    \\
Example 3&$(\theta,\alpha)$& $5.663\cdot10^{-30}$& $7.881\cdot10^{-31}$\\
&(31.265, 1)&$(11.313\sigma)$&$[11.301\sigma,11.326\sigma]$\\
 &  &   &    \\
\noalign{\global\arrayrulewidth0.05cm}
 \hline
 \noalign{\global\arrayrulewidth0.05pt}

\end{tabular}
\vspace{0.1cm}
\caption{TOHM p-values computed via \eqref{replacement1} and $\sigma$-significance. \textcolor{black}{The  last column refers to the Monte Carlo errors associated with the estimates of $\widehat{E[\phi({\mathcal{A}}_{c_k})]}$ that are used in \eqref{system} to obtain \eqref{replacement1} as described in the Section \ref{graphs}}.}
\label{real_table}}
\end{table} 
\textcolor{black}{
In   Figure \ref{assess}  we show  as red dashed lines the Monte Carlo estimates of  $P(\sup_{\bm{\theta} \in \bm{\Theta}}\{W(\bm{\theta})\}>c)$ obtained by simulating  data  under the null model for each of  example; the  Monte Carlo errors are given by the pink areas. For Examples 1 and 3, we simulated $130,000$ datasets, each of size $10,000$. For Example 2, a sample size of $100,000$ was needed in order to guarantee the marginal $\bar{\chi}^2_{01}$ asymptotic distribution of the LRT statistics. This, along with the three dimensional constrained optimization needed  to compute the LRT for each $\bm{\theta}$ at each replicate,   drastically reduced the computational speed. Therefore, for this example we only considered a simulation of $10,000$ Monte Carlo replicates.
The Monte Carlo estimates of $P(\sup_{\bm{\theta} \in \bm{\Theta}}\{W(\bm{\theta})\}>c)$} are compared with  the approximation in \eqref{replacement1} plotted as blue dashed lines as $c$ increases ($x$-axis). Specifically, \eqref{replacement1} has been computed via a set of $100$ Monte Carlo replicates \textcolor{black}{(each of size $10,000$ for Examples 1 and 3 and $100,000$ for Example 2)},  to estimate the quantities $E[\phi(\mathcal{A}_{c_k})]$   in \eqref{system}, and   the Lipschitz-Killing curvatures $\mathcal{L}^*_d(\bm{\Theta})$ in 
\eqref{expect2}.

For Example 1 (left panel of Figure \ref{assess}), we consider   a grid of size $R=2821$ over the $30$ degree radius circular search region centered at ($195$ RA,$28$ DEC). Since in this case $\bm{\Theta}$ is given by a disc, its EC is one and thus $\mathcal{L}_0(\bm{\Theta})=1$. In order to estimate $\mathcal{L}_1(\bm{\Theta})$ and $\mathcal{L}_2(\bm{\Theta})$ we consider $c_1=1$ and $c_2=8$, which lead to $\widehat{\mathcal{L}^*_1(\bm{\Theta}) }=-244.053$ and $\widehat{\mathcal{L}^*_2(\bm{\Theta})}= 644.244$. \textcolor{black}{As shown in the left panel of Figure \ref{assess}, the approximation becomes particularly accurate as $c$ approaches $24$; this corresponds to the threshold required for a $3\sigma$ detection (see \eqref{sigma} for a definition of ``$\sigma$-significance'').}

For Example 2 in the central panel of Figure \ref{assess}, we define  a grid of size $R=2500$ over the square $[1,10]\times[0.2,5]$. Again $\mathcal{L}_0(\bm{\Theta})=1$, and we chose $c_1=2$ and $c_2=3$. The resulting estimates for the Lipschitz-Killing curvatures are $\widehat{\mathcal{L}^*_1(\bm{\Theta}) }=30.11037$ and $\widehat{\mathcal{L}^*_2(\bm{\Theta})}= 30.52665 $. \textcolor{black}{In this case, despite the small simulation size,  a good approximation of $P(\sup_{\bm{\theta} \in \bm{\Theta}}\{W(\bm{\theta})\}>c)$ is quickly achieved as $c$ approaches 10.}

Finally for Example 3,  the parameter space $ \bm{\Theta}$ corresponds to $[-12,12]\times\{0,1,2\}$, and  we let $R=150$ as  we only allow values of $\alpha$ equal to $0,1$ and $2$. Selecting $c_1=0.5$ and $c_2=1$, we obtain $\widehat{\mathcal{L}^*_1(\bm{\Theta})} =16.724$ and $\widehat{\mathcal{L}^*_2(\bm{\Theta})}= 23.291$.
Despite the lack of smoothness in the underlying process,  \eqref{replacement1} leads to an upper bound for the global p-value, see the right panel of Figure \ref{assess}. 
\textcolor{black}{We conjecture that the fact that we obtain an upper bound on the p-value rather than a close approximation is due to the  several jumps in the derivative of $\{W(\bm{\theta})\}$ with respect to $\theta$. This is despite the fact that we only consider three possible values of $\alpha$. Thus, the resulting excursion set only involves simple connected components and no holes. Therefore, in this setting, $\widehat{E[\phi({\mathcal{A}}_{c_k})]}$ corresponds to the expected number of local maxima and, as shown in \eqref{localMax}, provides a bound on the global p-value, $P(\sup_{\bm{\theta} \in \bm{\Theta}}\{W(\bm{\theta})\}>c)$, from above. }

\textbf{\emph{Guidelines for setting  $c_k$.}} 
\textcolor{black}{  
Our goal is to reduce the computational time needed to compute the right hand side of \eqref{technical_bound3}
while guaranteeing  that the difference between the right and left hand sides of \eqref{technical_bound3} approaches zero. Since \eqref{expect2} holds for any choice of $c_k$, $k=1,\dots,D$, this can be done by selecting the thresholds  $c_k$ sufficiently small so that the  excursion sets $\mathcal{A}_{c_k}$ are composed by a reasonably high number of connected components. This reduces the size of the  Monte Carlo simulation needed to accurately estimate the quantities $E[\phi({\mathcal{A}}_{c_k})]$.}  Hence,  the $c_k$ should be chosen to be small enough that  $\mathcal{A}_{c_k}$ is  non-empty with high probability.
Additionally, since both the size and the sparsity of the graph $\mathcal{G}^D_k$ affect the running time of Algorithm \ref{algo}, $c_k$ should be selected accordingly.
These points can be assessed with a sensitivity analysis. Specifically, for a given $c_k$, $\mathcal{G}^D_k$ allows a two-dimensional visualization of the  $D$-dimensional mesh $\mathcal{M}_k$, and thus after step 2 in Algorithm \ref{algo}, $c_k$ can be increased to increase sparsity and decrease the size of  $\mathcal{G}^D_k$ before proceeding with steps 3-5.

\textcolor{black}{In principle,  the choice of  $c_1,\dots,c_D$ should also take into account the possibility that the ECs computed at  different thresholds $c_k$, $\phi({\mathcal{A}_{c_k}})$, may  be positively correlated leading to  inflation of the variance of the estimators of $E[\phi({\mathcal{A}}_{c_k})]$.} However, since we are interested in the limit as $c\rightarrow\infty$ and the Monte Carlo error associated with \eqref{replacement1} become extremely small as $c$ increases, such correlation may be of little concern. This may be true even when, as in  Figure \ref{assess}, the quantities $\phi({\mathcal{A}_{c_k}})$ have been computed on the same set of  Monte Carlo simulations for each $c_k$ considered.

\subsection{Data analysis}
We calculated  the TOHM p-value in \eqref{replacement1}  for  the case studies introduced in Section \ref{casestudies}.  The results are summarized in Table \ref{real_table}. In addition to the p-values, we report the respective $\sigma$-significance, a quantity  typically used in physics to quantify the statistical evidence in support of new discoveries, i.e., 
\begin{equation}
\label{sigma}
\#\sigma=\Phi^{-1}(1-\text{p-value}),
\end{equation}
where $\Phi$ is the standard normal cumulative function.

 In Example 1, we performed $R=2821$ tests over our circular search region centered at ($195$ RA, $28$ DEC). In our realistic simulation, the true dark matter emission was located at ($174.952$ RA, $37.986$ DEC) and the LRT-process used in TOHM achieves its maximum at ${\bm{\theta}}=$($175$ RA, $38$ DEC) with about $10\sigma$ significance. Notice that our original dataset includes 51,098 background events and only 39 dark matter events; hence the procedure appears to be particularly powerful even in presence of a low signal-to-noise ratio.
The   location at which the maximum LRT statistics has been observed  is plotted as a white circle   in the upper panel of Figure \ref{real_plots}.

In Example 2, we set $R=2500$ when testing \eqref{test1}  and the gamma model is rejected at a $0.05$ significance level by the THOM p-value. Whereas, when testing \eqref{eta1}, the log-normal model cannot be rejected; the resulting p-value is $0.596$. Thus, the  log-normal model  is selected for the maize seeds strength data, and the LRT-process achieves its maximum at ${\mu}=5.004$ and ${\sigma}=0.633$. The log-normal fitted model is plotted in the bottom left panel of Figure \ref{real_plots} as a red solid line.

 Finally in Example 3, testing \eqref{test2} $R=150$ times, \eqref{replacement1}   provides strong evidence ($\sim 11\sigma$) in favor of a linear trend  ($\alpha=1$) with a break point at $\theta=31.265$. Hence we expect the risk of giving birth to a child with down syndrome to increase when the mother is 31 years old or older. The  model selected  is displayed as a red  solid line in the bottom right panel of  Figure \ref{real_plots}, with the break-point indicated by a red triangle. For the sake of comparison, we also plot the fitted model when allowing a quadratic trend ($\alpha=2$) with a break point choosen at $\hat{\theta}=20$.

\section{Discussion}
\label{discussion}  

In this paper we propose a novel computational method to perform TOHM in the  multidimensional setting. The resulting inferential tool generalizes classical inferential methods,  such as the Likelihood Ratio Test, beyond standard regularity conditions including  non-identifiability of multidimensional parameters and  non-nestedness of the models under comparison.
From a more practical perspective, the  procedure proposed provides a computationally efficient solution to the bump hunting problem in multiple dimensions, and implicitly introduces a type I error correction for dependent tests. It also simplifies the estimation of the so called Lipschitz-Killing curvatures involved in  the computation of the TOHM p-value on the basis of \citet{taylor2008}.

Despite its simplicity and efficiency in computation, the main limitation of TOHM is that it requires the specification of  a parametric form  for the alternative model. In the context of signal identification for instance, this implies that the researcher can specify  the density function of the events associated to the signal (e.g, a Gaussian bump). In situations where this cannot be done, one possibility is to refer to nonparametric inferential methods \citep[e.g.,][]{chen2016, deep_mode, bkg}. \textcolor{black}{More work is needed to extend TOHM to discrete regions $\bm{\Theta}$ and provide a formal justification of its validity in non-regular setting, such as the one in  Example 3.}

It is important to note that, in the context of multiple hypothesis testing and large-scale inference,  TOHM allows us  to reduce the dimensionality of the tests being conducted  from $R$ to one by exploring the topology of the random field associated with the test statistics of interest. From this perspective, TOHM may offer a path forward to solve the long-standing problem of  identifying an unknown number  of  signals, in one or multiple dimensions. 
\appendix
\textcolor{black}{
\section{Regularity conditions and proofs}
\label{regular}
\paragraph{Regularity conditions A0-A5.}
\citet{ghosh} show that  when testing \eqref{test1} for \eqref{mixture} with $\bm{y}\in \Real$ and $\bm{\Theta}\subset \Real$, the LRT converges to a $\bar{\chi}_{01}^2$ process under suitable assumptions. In our examples the nuisance parameter $\bm{\theta}$ is allowed to be multidimensional, i.e., $\bm{\Theta}\subset \Real^D$, $D\geq 1$, and the data can be multivariate, i.e., $\bm{y}\in \Real^q$, $q\geq 1$; therefore, we re-state the regularity conditions of \citet{ghosh} accordingly below. 
\begin{enumerate}
\item[A0 - ] The mixture model in \eqref{mixture} is strongly identifiable, i.e., 
\[
\text{if $\eta\in(0,1)$ and }h(\bm{y},\eta,\bm{\gamma},\bm{\theta})=h(\bm{y},\eta',\bm{\gamma}',\bm{\theta}')\quad\Rightarrow\quad (\eta,\bm{\gamma},\bm{\theta})=(\eta',\bm{\gamma}',\bm{\theta}'). \]
\item[A1 - ] For each  $\bm{\theta}$ fixed, let $S(\bm{\theta})= \nabla \log h(\bm{y},\eta,\bm{\gamma},\bm{\theta})$ be the score vector and $S_{j}(\bm{\theta})$ its element $j$. Denote them with  $S^0(\bm{\theta})$ and $S^0_{j}(\bm{\theta})$  when 
 evaluated at the true values of $(\eta,\bm{\gamma})$ under $H_0$, i.e., $(0,\bm{\gamma}_0)$. \\ Similarly, for each  $\bm{\theta}$ fixed, let $\bm{H}(\bm{\theta})$ be the Hessian matrix of $\log h(\bm{y},\eta,\bm{\gamma},\bm{\theta})$ and $H_{jk}(\bm{\theta})$ its element in position $(j,k)$. Denote them with $\bm{H}^0(\bm{\theta})$ and  $H^0_{jk}(\bm{\theta})$ when evaluated at  $(\eta,\bm{\gamma})=(0,\bm{\gamma}_0)$. 
We require the following.
\begin{enumerate}
\item[(i)] $\bm{\Gamma}$ is an open interval in $\Real^p$ and $\Theta$ is a compact subset of $\Real^D$.
\item[(ii)] $h(\bm{y},\eta,\bm{\gamma},\bm{\theta})$ is continuous in $(\eta,\bm{\gamma},\bm{\theta})$ and twice continuously differentiable with respect to $(\eta,\bm{\gamma})$.
\item[(iii)]  $E[S^0_j(\bm{\theta})]=0$ for all $j=1,\dots,p+1$ and $\bm{\theta}\in \bm{\Theta}$.
\item[(iv)] $E[H^0_{jk}(\bm{\theta})]=-E[S^0_j(\bm{\theta})S^0_k(\bm{\theta})]$ for all $j,k=1,\dots,p+1$  and $\bm{\theta}\in \bm{\Theta}$.
\item[(v)] For $j,k=1,\dots,p+1$,
\[\lim_{\delta\rightarrow 0}E[\sup_{\substack{||(\eta,\bm{\gamma},\bm{\theta})-(0,\bm{\gamma}_0,\bm{\theta})||<\delta}}
 |H_{jk}(\bm{\theta})-H^0_{jk}(\bm{\theta})|]= 0.\]
\end{enumerate}
Where all the expectations above and those to follow are taken with respect to $f(\bm{y},\bm{\gamma}_0)$.
\item[A2 - ] There exists a compact neighborhood $\mathcal{N}$ of $(\eta,\bm{\gamma})$ such that 
$E[\psi(\bm{y},\bm{\theta})]<0$, where 
\[\psi(\bm{y},\bm{\theta})=\sup_{(\eta,\bm{\gamma})\in[0,1]\times\mathcal{N}^c}\log\frac{h(\bm{y},\eta,\bm{\gamma},\bm{\theta})}{f(\bm{y},\bm{\gamma})}.\]
Further $\psi(\bm{y},\bm{\theta})$ in continuous on $\bm{\Theta}$ and there exist a function $w(\bm{y})$ such that $|\psi(\bm{y},\bm{\theta})|\leq w(\bm{y})$ and $E[w(\bm{y})]\leq\infty$ for all $\bm{\theta}\in \bm{\Theta}$.
\item[A3 - ] For each $(\eta,\bm{\gamma})\neq(0,\bm{\gamma}_0)$ there exists an open ball with center at $(\eta,\bm{\gamma})$ and radius $\delta_0$, namely $B(\delta_0)$, such that
\[|\psi_B(\bm{y},\bm{\theta})|\leq w(\bm{y})\quad\text{for all $\bm{\theta}\in \bm{\Theta}$}\]
where
\[\psi_B(\bm{y},\bm{\theta})=\sup_{(\eta',\bm{\gamma}')\in B(\delta_0)\cap[0,1]\times\bm{\Gamma}}\log \frac{h(\bm{y},\eta',\bm{\gamma}',\bm{\theta})}{f(\bm{y},\bm{\gamma}_0)} \qquad \text{and } \qquad E[w(\bm{y})]\leq\infty.\]
\item[A4 - ] $\bm{I}(\bm{\theta})=E[-\bm{H}^0(\bm{\theta})]$ is continuous in $\bm{\theta}$ and positive definite uniformly over $\bm{\Theta}$.
\item[A5 - ] $E \biggl| \frac{g(\bm{y},\bm{\theta})}{f(\bm{y},\bm{\gamma}_0)}- \frac{g(\bm{y},\bm{\theta}^\dag)}{f(\bm{y},\bm{\gamma}_0)}\biggl|^\xi \leq K ||\bm{\theta}-\bm{\theta}^\dag||^{1+\lambda}$ for some $\xi,\lambda>0$ and $K\geq 0$.
\end{enumerate}}

\textcolor{black}{
\paragraph{Proof of Proposition \ref{prop1}.}
Under conditions A0-A1, it follows from  \citet[equations $(2.7)$ and $(2.8)$]{ghosh} that the random field $\{W_n(\bm{\theta})\}$ can be written as:
\begin{equation}
\label{reminder}
\begin{split}
\{W_n(\bm{\theta})\}&=\{R_n(\bm{\theta})\} \quad \text{over the set $\bm{\Theta}_{0n}=\{\bm{\theta}:T_n(\bm{\theta})<0\}$}\\
\{W_n(\bm{\theta})\}&=\{T_n(\bm{\theta})\}+ \{R_n(\bm{\theta})\} \quad \text{over the set $\bm{\Theta}_{1n}=\{\bm{\theta}:T_n(\bm{\theta})\geq0\}$}\\
\end{split}
\end{equation}
where $\{R_n(\bm{\theta})\}$ is a  reminder term such that, if A2-A3 hold, $\{R_n(\bm{\theta})\}=o_p(1)$, uniformly in $\bm{\theta}$ . Whereas, 
$\{T_n(\bm{\theta})\}$ is a random field such that, if A4 and A5 hold and under $H_0$, it converges weakly to a  Gaussian random field $\{Z(\bm{\theta})\}$ with mean zero, unit variance and covariance function depending on $\bm{\theta}$.}

\textcolor{black}{
\paragraph{Proof of Proposition \ref{prop2}.}
Let $\phi_{10}$ and $\phi_{20}$ be the true values of $\phi_{1}$ and $\phi_{2}$ in \eqref{logistic} when $H_0$ in \eqref{test2} is true.
Denote with  $U^\star_n(\bm{\theta})$  the normalized score function of \eqref{logistic}  for $\bm{\theta}=(\theta,\alpha)$ fixed. Under $H_0$, $U^\star_n(\bm{\theta})$ can be specified by 
\begin{equation}
\label{score}
U^\star_n(\bm{\theta}|H_0)=\sum_{i=1}^n\frac{Z_i}{\sigma}\sqrt{m_i\pi_{0i}(1-\pi_{0i})}(x_i-\theta)^{\alpha}\mathbbm{1}_{\{ x_i\geq\theta \}}
\end{equation} 
where $\sigma=\sum_{i=1}^n m_i\pi_{0i}(1-\pi_{0i})(x_i-\theta)^{2\alpha}\mathbbm{1}_{\{ x_i\geq\theta \}}$, $Z_i=\frac{Y_i-m_i\pi_{0i}}{\sqrt{m_i\pi_{0i}(1-\pi_{0i})}}$, $Y_i\sim \text{Binomial}(m_i,\pi_{i})$,  $\pi_i=\bigl[1+\exp{\{-\phi_1+\phi_2 x_i+\xi(x_i-\theta)^{\alpha}\mathbbm{1}_{\{ x_i\geq\theta \}}\}}\bigl]^{-1}$ and such that, under $H_0$, $\pi_i=\pi_{0i}=\bigl[1+\exp{\{-{\phi}_{10}+{\phi}_{20} x_i\}}\bigl]^{-1}$.   Under the Cramer's classical  conditions \citep[][p.500]{cramer2}, $U^\star_n(\bm{\theta}|H_0)$  is asymptotically normally distributed with mean zero and variance one. Since $\phi_{10}$ and $\phi_{20}$ are unknown, we can estimate them by the respective MLEs $\widehat{\phi}_{10}$ and $\widehat{\phi}_{20}$.  The latter are $\sqrt{n}$-consistent estimators of $\phi_{10}$ and $\phi_{20}$, therefore, when substituting $\widehat{\pi}_{0i}=\bigl[1+\exp{\{-\widehat{\phi}_{10}+\widehat{\phi}_{20} x_i\}}\bigl]^{-1}$  to $\pi_{0i}$ in \eqref{score}, the Gaussian asymptotic distribution of $U^\star_n(\bm{\theta}|H_0)$ is preserved. 
By De Moivre-Laplace theorem, under $H_0$, each $Z_i$ follows  a standard normal distribution as $m_i\rightarrow\infty$.
Therefore, for each $\bm{\theta}$ fixed,    $U^\star_n(\bm{\theta}|H_0)$ is a linear function of the $Z_i$; hence the joint distribution of $\{U^\star_n(\bm{\theta}|H_0)\}$ is  asymptotically   Gaussian as $m_i\rightarrow\infty$ for all $i=1,\dots,n$. Finally, by virtue of the equivalence between  $U^\star_n(\bm{\theta})$ and  $W_n(\bm{\theta})$ \citep{moran70,davies77}, $\{W_n(\bm{\theta})\}$ is also  asymptotically distributed as a   Gaussian random field   under $H_0$. 
}

\section{EC densities for Guassian, $\chi^2$ and $\bar{\chi}_{01}^2$ random fields}
\label{ECapp}
\textbf{Gaussian case.} If $\{W(\bm{\theta})\}$ is such that $W(\bm{\theta})\sim N(0,1)$ for all $\bm{\theta}$, the EC densities $\rho_d(c)$, $d=0,\dots,5$ are given by
\begin{equation*}
\begin{matrix}
\rho_0(c)=1-\Phi(c), \qquad& \rho_1(c)= \frac{e^{-\frac{c^2}{2}}}{2\pi},\qquad&\rho_2(c)=  \frac{e^{-\frac{c^2}{2}}}{(2\pi)^{3/2}},\\
\rho_3(c)= \frac{(c^2-1)e^{-\frac{c^2}{2}}}{(2\pi)^{2}}, \qquad& \rho_4(c)=  \frac{(c^3-3c)e^{-\frac{c^2}{2}}}{(2\pi)^{5/2}}, \qquad& \rho_5(c)= \frac{(c^4-4c^2+3)e^{-\frac{c^2}{2}}}{(2\pi)^{3}}, \\
\end{matrix}
\end{equation*}
where $\Phi(\cdot)$ is the cumulative function of a standard normal.

\textbf{$\chi_s^2$ case.} If $\{W(\bm{\theta})\}$ is such that $W(\bm{\theta})\sim \chi_s^2$ for all $\bm{\theta}$, the EC densities $\rho_d(c)$, $d=0,\dots,3$ are given by
\begin{equation*}
\begin{matrix}
\rho_0(c)=1-F_\chi(c), \qquad& \rho_1(c)= \frac{c^\frac{s-1}{2}}{\Gamma(\frac{s}{2})}\sqrt{\frac{2}{\pi}}e^{-\frac{c}{2}},\qquad&\rho_2(c)=\bigl(\frac{c}{2}\bigl)^{\frac{s}{2}-1}\frac{e^{-\frac{c}{2}}}{2\pi}\bigl[c-(s-1)\mathbb{1}_{\{s\geq2\}}\bigl], \\
\end{matrix}
\end{equation*}
\begin{equation*}
\begin{split}
\rho_3(c)=&\frac{c^{\frac{s-3}{2}}e^{-\frac{c}{2}}}{(2\pi)^{3/2}\Gamma\bigl(\frac{s}{2}\bigl)2^{\frac{s-2}{2}}}\bigl[(s-1)(s-2)\mathbb{1}_{\{s\geq3\}}-2(s-1)c\mathbb{1}_{\{s\geq2\}}+(c^2-c)\mathbb{1}_{\{s\geq1\}}\bigl],\\
\end{split}
\end{equation*}
where $F_\chi(\cdot)$ is the cumulative function of a $\chi_s^2$ and $\mathbb{1}_{\{\cdot\}}$ is the indicator function. 

\textcolor{black}{
\textbf{$\bar{\chi}_{01}^2$ case.}
From \citet{taylor13} it follows that the  EC densities of a $\bar{\chi}_{01}^2$ random field are given by the sum of the EC densities  of a $\chi^2_{0}$  random field (i.e., a random field which is zero everywhere) and those of a $\chi^2_1$ random field, each multiplied by the respective   weight, i.e., 0.5. }
Consequently,  when $\bm{\Theta} \subset \Real^2$ as in Examples 1 and 2, \eqref{ECpval} specifies as
\begin{equation}
\label{DTKmix}
E[\phi(\mathcal{A}_c)]=\frac{c^{\frac{1}{2}}e^{-\frac{c}{2}}}{(2\pi)^{\frac{3}{2}}}\mathcal{L}_2(\bm{\Theta})+\frac{e^{-\frac{c}{2}}}{2\pi}\mathcal{L}_1(\bm{\Theta})+\frac{P(\chi^2_1>c)}{2}\mathcal{L}_0(\bm{\Theta})
\end{equation}
where the functions of $c$ multiplying $\mathcal{L}_0(\bm{\Theta}),\dots,\mathcal{L}_2(\bm{\Theta})$ are the EC densities of a two-dimensional $\chi_1^2$ random field  divided by 2. Because the EC densities of a two-dimensional $\chi_0^2$ random field evaluated at  $c>0$ are all zero, they do not contribute in \eqref{DTKmix}.

See \citet[p.426]{adlerbook} for higher order EC densities. 

\section*{Acknowledgements}{
\textcolor{black}{The authors thank two anonymous referees and the associate editor, whose comments and suggestions greatly improved the quality and clarity of the manuscript}. SA and DvD also thank
Jan Conrad for the valuable discussion of the physics problems which motivated this work, and  Brandon Anderson who provided the Fermi-LAT datasets used in the analyses. SA acknowledges support from the Swedish Research Council through a grant with PI: Jan Conrad.
 DvD acknowledges support from the Marie-Skodowska-Curie RISE (H2020-MSCA-RISE-2015-691164) Grant provided by the European Commission. 
 }

\section*{Supplementary Material}{
The folder \texttt{\emph{Codes\_JCGS}} collects the data used in Example 1, the \texttt{R} package \texttt{TOHM} and  the codes used  for the analyses  in Table \ref{real_table} and Figure \ref{real_plots}. \textcolor{black}{In the file  \texttt{\emph{Supp\_JCGS.pdf}}  we  assess the  validity of assumptions A0-A5 for Examples 1 and 2.  }}

\fontsize{9}{9}\selectfont{
\bibliography{biblioBio2}
}

\end{document}